\documentclass[fleqn]{vch-book}
\usepackage{amsmath}
\usepackage{amssymb}
\usepackage{makeidx}
\usepackage{graphicx}
\usepackage{times}
\usepackage{tabularx}
\usepackage{booktabs}
\usepackage{cite}
\begin{document}
\section{Energy spectra and eigenstates of quasiperiodic
tight-binding Hamiltonians}
\authorafterheading{Uwe Grimm}
\affil{Applied Mathematics Department,
Faculty of Mathematics and Computing,\\
The Open University, Walton Hall, 
Milton Keynes MK7 6AA, United Kingdom}
\authorafterheading{Michael Schreiber}
\affil{School of Engineering and Science, 
International University Bremen,\\
P.O.\ Box 750\thinspace 561, 28725 Bremen, Germany}

\subsection{Introduction}

Among the physical properties of quasicrystals \cite{GS01}, their
transport properties, such as the electric conductivity, have
attracted particular attention. Over the past decades, a lot of
experimental and theoretical effort has been spent on the
investigation and explanation of transport anomalies in quasicrystals,
see \cite{S99,BM00,SSH02} for recent collections of review articles
and \cite{BG02} for further references; compare also
\cite{KSB,KLS,SL}. Though there undoubtedly exists a strong
interrelation between the quasicrystalline structure and the
electronic properties \cite{HHBKL,KWB} --- after all, the electrons
determine the structure of a solid --- it makes sense to consider the
properties of electrons in a given aperiodic solid.

The simplest model that may be expected to capture at least some
characteristics of electronic properties of quasicrystals is that of a
single electron moving in a quasiperiodic background, realised, for
instance, by a quasiperiodic potential. Mathematical and theoretical
studies of related systems had already been performed prior to the
experimental discovery of quasicrystals, see, for instance, the review
\cite{S82} on the properties of almost periodic Schr\"{o}dinger
operators. In most cases, the models are defined on a discrete space,
physically motivated by a tight-binding approach where the electron
can hop from one vertex of a graph to any of its neighbouring
vertices, and the aperiodicity is introduced either by an aperiodic
graph to model the aperiodic solid, or by an aperiodic modulation of
the potential or the hopping probabilities.

While quite a lot is known rigorously about one-dimensional systems,
see \cite{D00} and references therein, higher-dimensional systems,
apart from a few particular examples, have mainly been investigated by
numerical calculations for finite systems or periodic
approximants. Here, we summarise several results for aperiodic
tight-binding models in two and three dimensions, which are mainly
based on numerical investigations. These concern energy spectra of
quasiperiodic tight-binding models and the corresponding energy level
statistics, multifractal properties of the eigenstates, and quantum
diffusion. In addition, we present some recent results obtained for
interacting electrons in one-dimensional systems.

We consider the simplest tight-binding model of a single
electron\index{electrons!non-interacting} moving on a (quasiperiodic)
graph, for instance the graph corresponding to the rhombic Penrose
tiling. The Hamiltonian has the form $H_{jk} = t_{jk} +
\varepsilon_{k}\delta_{jk}$, where $j$ and $k$ label the vertices of
the graph. In a Dirac bra-and-ket notation with mutually orthogonal
states $|j\rangle$ associated to any vertex $j$, we
have\index{tight-binding model}
\begin{equation}
H =  \sum_{jk}|j\rangle t_{jk}\langle k| +
\sum_{k}|k\rangle\varepsilon_{k}\langle k| \label{eq:h}
\end{equation}
with matrix elements $H_{jk}=\langle j|H|k\rangle$. Here, $t_{jk}$
play the role of hopping elements between the states associated to
vertices $j$ and $k$, and will usually considered to vanish except
when the two vertices are connected by an edge (bond) of the
underlying quasiperiodic graph. In the simplest scenario, the hopping
elements are just $t_{jk}=1$ for vertices connected by an edge and
$t_{jk}=0$ otherwise. The parameters $\varepsilon_{k}$ correspond to
on-site energies. As a further simplification, we shall usually choose
$\varepsilon_{k}=0$ for all vertices $k$, so no vertex is
energetically preferred to any other.

In the simplest case, where the hopping elements $t_{jk}$ are either
one or zero and where the on-site energies vanish, the Hamiltonian is
just the adjacency matrix of the underlying graph, encoding which
vertices are neighbours connected by edges. For an infinite
quasiperiodic tiling the matrix will be infinite; in practice, we
either consider finite patches and try to extrapolate the results to
infinite systems, or investigate a series of periodic approximants,
i.e., periodic systems with growing unit cells which approximate the
infinite quasiperiodic tiling.

In any case, we are interested in the behaviour of the eigenvalues and
the corresponding eigenvectors of these matrices, in particular in the
infinite-size limit where the aperiodic system is approached. The
eigenvalues and eigenvectors are interpreted as the single-electron
energies and the corresponding wave functions.  The structure of the
density of states (DOS) as a function of the energy, which is defined
as the limit of the number of eigenvalues in an energy interval when
the size of the interval goes to zero, and the localisation properties
of the wave functions are intimately linked to the electronic
transport properties.

Quasiperiodic systems are very peculiar in this respect, due to the
competition between the aperiodicity on the one hand and the strict
quasiperiodic order on the other hand. Aperiodicity means that there
is no translation that maps the system into itself, so the system
eventually looks different from any of its vertices, even if local
configurations may be the same. This variation acts similar to a
random disorder and thus favours localisation of wave functions.  The
quasiperiodic order is reflected in the repetitivity of the tilings,
which means that the same local neighbourhoods reappear again and
again, albeit not in a periodic fashion. For the examples at hand,
this can in fact be phrased more strongly by giving bounds on the
distance between appearances of the same patches in the
tiling. Repetitivity causes resonances between equivalent local
configurations, and thus favours extended wave functions. The result
is that wave functions in quasiperiodic tight-binding models are
expected to be different from the exponentially localised wave
functions found in (strongly) disordered systems, like in the Anderson
model of localisation \cite{Ande58}, but also different from the Bloch
waves found in the periodic situation of a usual crystalline
system. Such wave functions are often called ``critical'', because
they appear at the metal-insulator transition in the three-dimensional
Anderson model of localisation \cite{Schr85,SG91,GS92,GS95}.

The Anderson model of localisation\index{localisation} is defined by a
single-particle Hamiltonian like (\ref{eq:h}) on a cubic lattice with
a nearest-neighbour hopping term and an on-site energetic
disorder. For this simple model, one finds a transition from a
metallic phase at small disorder to an insulating phase at large
disorder, corresponding to a transition from extended to exponentially
localised eigenstates \cite{HS93a,HS93b,HS94}.  At the transition
itself, the eigenstates become critical; they are neither extended nor
exponentially localised, but show a multifractal distribution of
amplitudes \cite{SG91}. In fact, the expectation that the wave
functions in quasiperiodic model systems are neither exponentially
localised nor extended Bloch-like states, has been substantiated by
rigorous arguments for large classes of one-dimensional discrete
aperiodic Schr\"{o}dinger operators constructed from substitution
rules, where it can be shown that their spectrum is purely singular
continuous\index{spectrum!singular continuous} \cite{D00}, and by
numerous numerical investigations of two- and three-dimensional
systems.

\subsection{Energy Spectra and Eigenstates}\index{spectrum}

We start by having a look at the spectrum of tight-binding
Hamiltonians of this kind. Typical results for the DOS and the
integrated density of states (IDOS), which just counts the number of
eigenvalues up to a given energy $E$, are shown in
Fig.~\ref{fig:dosidos} for examples in one, two and there spatial
dimensions.\index{density of states}\index{integrated density of
states}\index{tight-binding model}

\begin{vchfigure}[b]
\centerline{\includegraphics[width=0.5\textwidth]{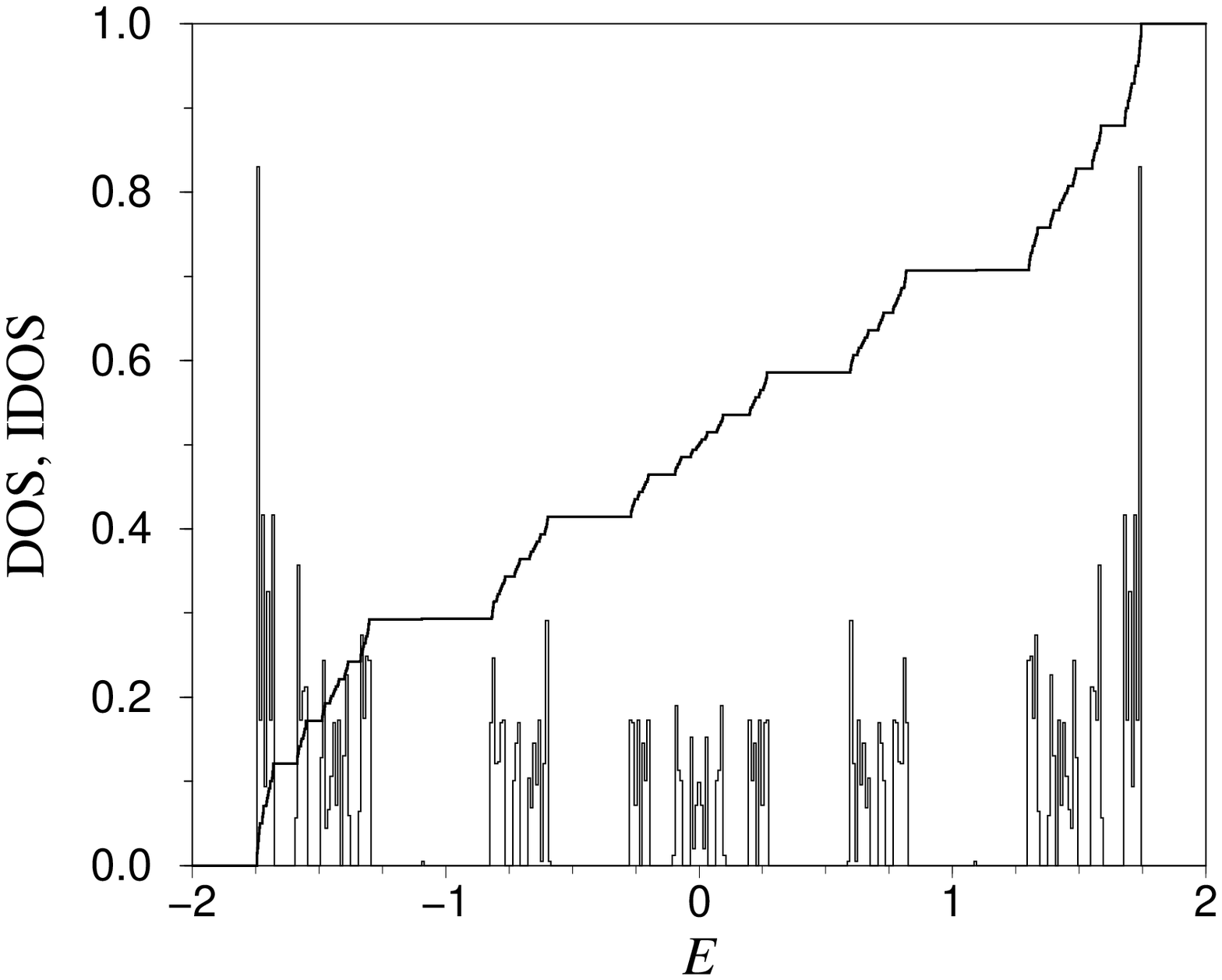}
\hspace*{\fill}
\includegraphics[width=0.5\textwidth]{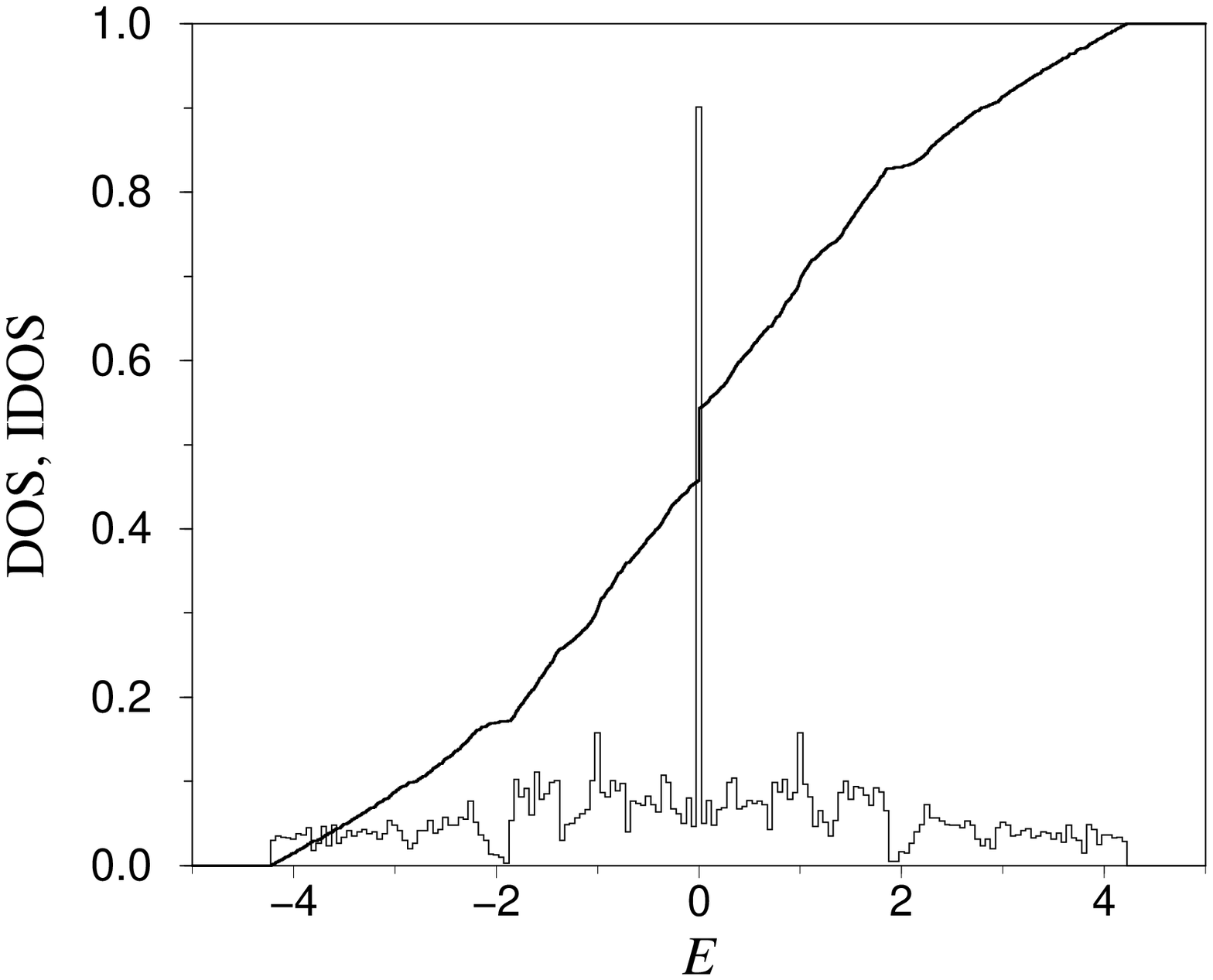}}
\centerline{\parbox[b]{0.5\textwidth}{%
\includegraphics[width=0.5\textwidth]{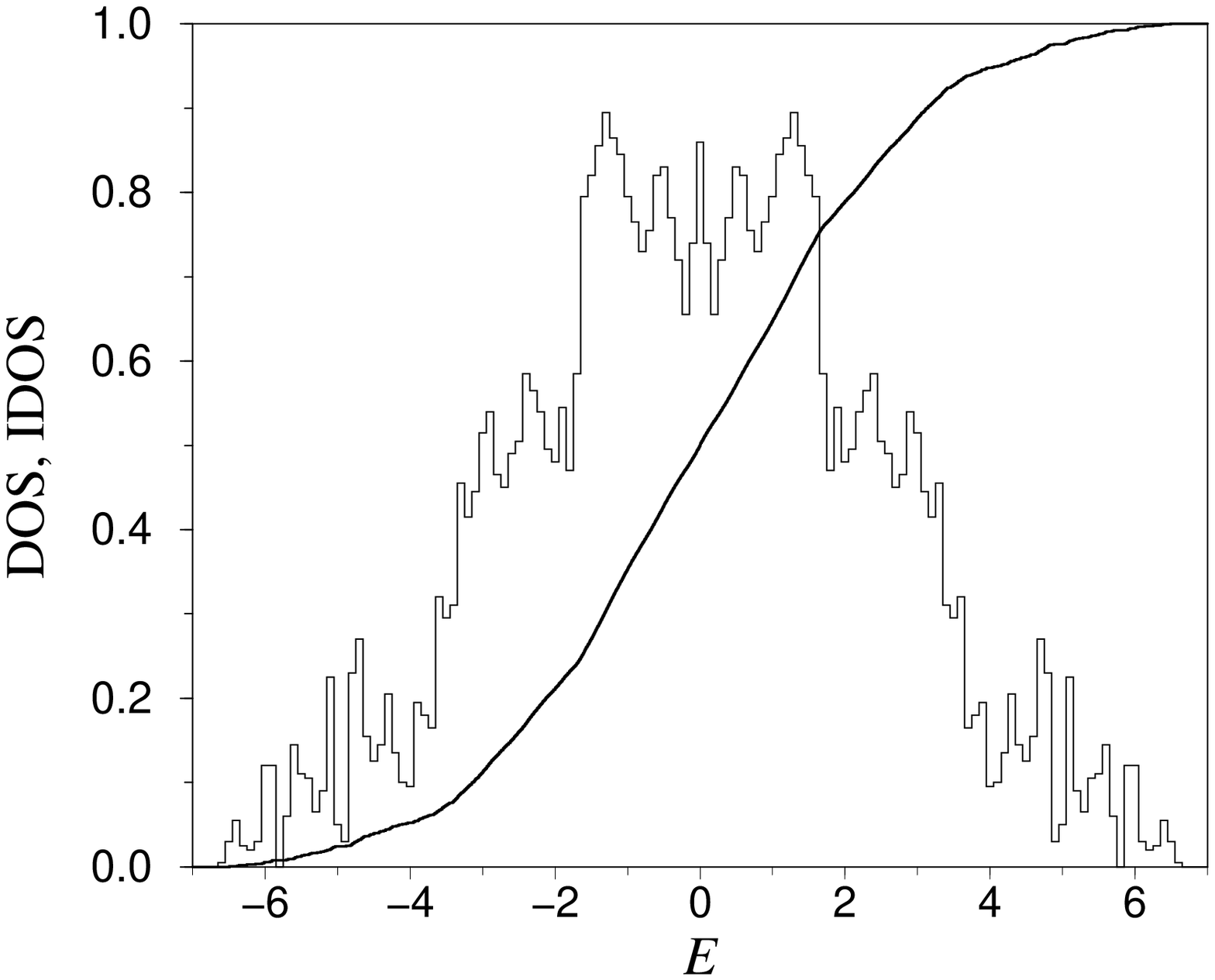}}%
\parbox[b]{0.5\textwidth}{\vchcaption{IDOS and DOS 
(arbitrary scale) of states of an octonacci
chain (top left), a periodic approximant of the octagonal
Ammann-Beenker tiling (top right) and of a periodic approximant 
of the icosahedral Ammann-Kramer-Neri tiling (bottom). The peak 
at $E=0$ for the planar case is due to energetically degenerate 
confined states.\bigskip}\label{fig:dosidos}}}
\end{vchfigure}

In the one-dimensional case, the example used is the co-called
octonacci chain \cite{YGRS00}. In one dimension, choosing all hopping
parameters as one and all on-site energies as zero yields the simple
periodic chain, so we have to introduce some parameter in order to
have a non-trivial aperiodic system. In this case, this is
conveniently achieved by assigning two values $t_{j,j+1}\in\{1,v\}$
for the hopping parameters such that the sequence of hopping
parameters along the chain is aperiodic; choosing an aperiodic
sequence for the on-site energies would lead to very similar
results. For the octonacci chain, the sequence is obtained from the
two-letter substitution rule\index{substitution rule}
\begin{equation}
\varrho:\;\begin{array}{lcl} S & \rightarrow & L \\
                             L & \rightarrow & LSL \end{array}
\label{eq:subst}
\end{equation}
applied repeatedly on the initial word $w_{0}=S$, so
$w_{m}=\varrho^{m}(w_{0})$. The limit word is semi-infinite aperiodic
sequence $w_{\infty}=LSLLLSLLSLLSLLLSL\ldots$.  We now choose the
hopping parameters as $t_{j,j+1}=1$ if the $j$th letter in
$w_{\infty}$ is an $L$, and as $t_{j,j+1}=v$ if it is an $S$. The
spectrum shown in Fig.~\ref{fig:dosidos} corresponds to a parameter
value $v=1/2$.

For such systems, it is known that the spectrum is purely singular
continuous,\index{spectrum!Cantor}\index{spectrum!measure} and is a
Cantor set of zero Lebesgue measure \cite{D00}. This is reflected in
the apparent set of gaps in the spectrum; the possible positions of
the corresponding plateaux in the IDOS are also known by Bellissard's
gap labelling theorem \cite{BBG92}. One powerful method to tackle
one-dimensional system employs the so-called trace maps, see
\cite{BGJ93,WGS00} and references therein for details.

The two-dimensional example in Fig.~\ref{fig:dosidos} corresponds to a
periodic approximant of the octagonal Ammann-Beenker
tiling\index{Ammann-Beenker tiling} \cite{AGS92,BGM02,BGM02e,GS02}; in
this case all hopping parameters are chosen as one along the edges and
zero otherwise. The same choice applies to the three-dimensional
system, which corresponds to a periodic approximant of the icosahedral
Ammann-Kramer-Neri tiling \cite{KN84}. There is a clear tendency of
smoothing of the spectrum as the dimension increases, and there are
only a few, if any, gaps in the spectrum. One particularity of the
two-dimensional case is the pronounced peak in the centre of the
spectrum at energy $E=0$, which is due to families of strictly
localised or ``confined'' states supported on a finite number of
vertices, which, due to the repetitivity of the tiling, reappear at
various places throughout the tiling and make up a finite fraction of
all states, see \cite{RS95} and references therein.  This is a
consequence of the local topology of the tiling, and also happens for
the tight-binding model on a rhombic Penrose tiling
\cite{RS95}.\index{states!critical}\index{states!extended}
\index{states!localised}\index{states!confined}

Some properties of such energy spectra will be discussed below in more
detail, including the level-spacing distribution and fractal
dimensions of the spectral measure. But first we are going to discuss
some general features of the eigenstates that are observed numerically
for two and three-dimensional tight-binding Hamiltonians defined on
quasiperiodic tilings.

Again, the one-dimensional situation has been studied in much detail,
see \cite{D00} and references therein for details. For many models
based on substitution sequences, such as our example of the octonacci
chain, it is known rigorously that the generalised eigenstates are
neither exponentially localised --- in which case they would
correspond to proper eigenvalues and hence yield a discrete
spectrum\index{spectrum!discrete} --- nor extended over the entire
system. These are the critical states mentioned previously. However,
this is definitely not true for all one-dimensional aperiodic discrete
Schr\"{o}dinger operators; a classical example is provided by the
Aubry-Andr\'{e} or Harper model,\index{Harper model} in the
mathematical literature also known as the almost-Mathieu equation, see
\cite{AA80,EGRS99} and references therein.  This model is also a
one-dimensional quasiperiodic tight-binding model, but now the local
on-site potential takes values in a continuous interval, in contrast
to the substitution models discussed above. The potential at position
$j$ of the chain, where $j$ is an integer, has the form
$V(j)=2\mu\cos(\alpha j+\beta)$, which is aperiodic provided
$\alpha/2\pi$ is irrational. This model behaves rather differently
from the substitution chains, and details may even depend on the type
of irrationality of $\alpha/2\pi$ --- irrational numbers that are too
well approximated by rational numbers may show a non-generic
behaviour. In most cases, the golden mean or its inverse is used
\cite{AA80}. Depending on the strength of the quasiperiodic potential,
one finds, for $\mu<1$, a metallic phase where {\em all}\/ eigenstates
are extended, and the spectrum is absolutely
continuous,\index{spectrum!continuous} and, for $\mu>1$, an insulating
phase, where {\em all}\/ eigenstates are exponentially localised, and
the spectrum is pure point.\index{spectrum!point} Here, the hopping
parameter was assumed to be unity. For the critical value
$\mu=\mu_{\rm c}=1$, one observes a metal-insulator transition with
multifractal eigenstates, similar to that found in the
three-dimensional Anderson model of localisation\index{Anderson model}
\cite{Ande58,SG91,GS92,GS95} and also similar to the behaviour of
substitution-based systems like the octonacci chain. In contrast to
the Anderson model, there are no mobility edges \cite{GS95} in the
Aubry-Andr\'{e} model --- at the value $\mu_{\rm c}=1$ the entire
spectrum localises. And, of course, there is no randomness whatsoever
in this model --- the metal-insulator transition takes place solely as
a consequence of the quasiperiodic potential strength. This model has
also been used as a one-dimensional toy model to study the effects of
an electron-electron interaction on the metal-insulator transition
\cite{EGRS99,ERS01,ERS02,SRS02a,SRS02b}, see Sec.~\ref{sec:ie}
below.\index{metal-insulator transition}

\begin{vchfigure}[b]
\centerline{\includegraphics[width=0.48\textwidth]{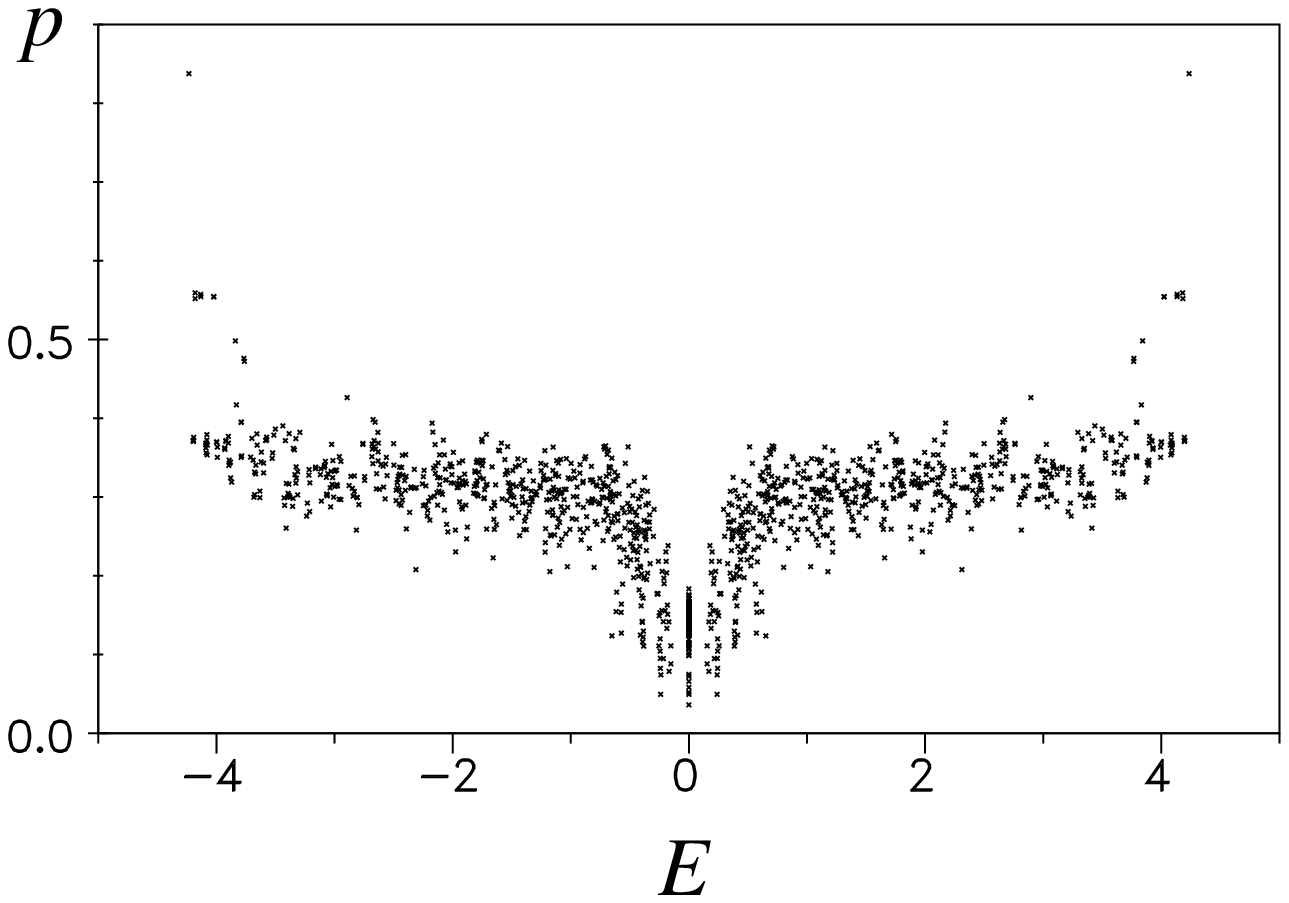}
\hspace*{\fill}\includegraphics[width=0.48\textwidth]{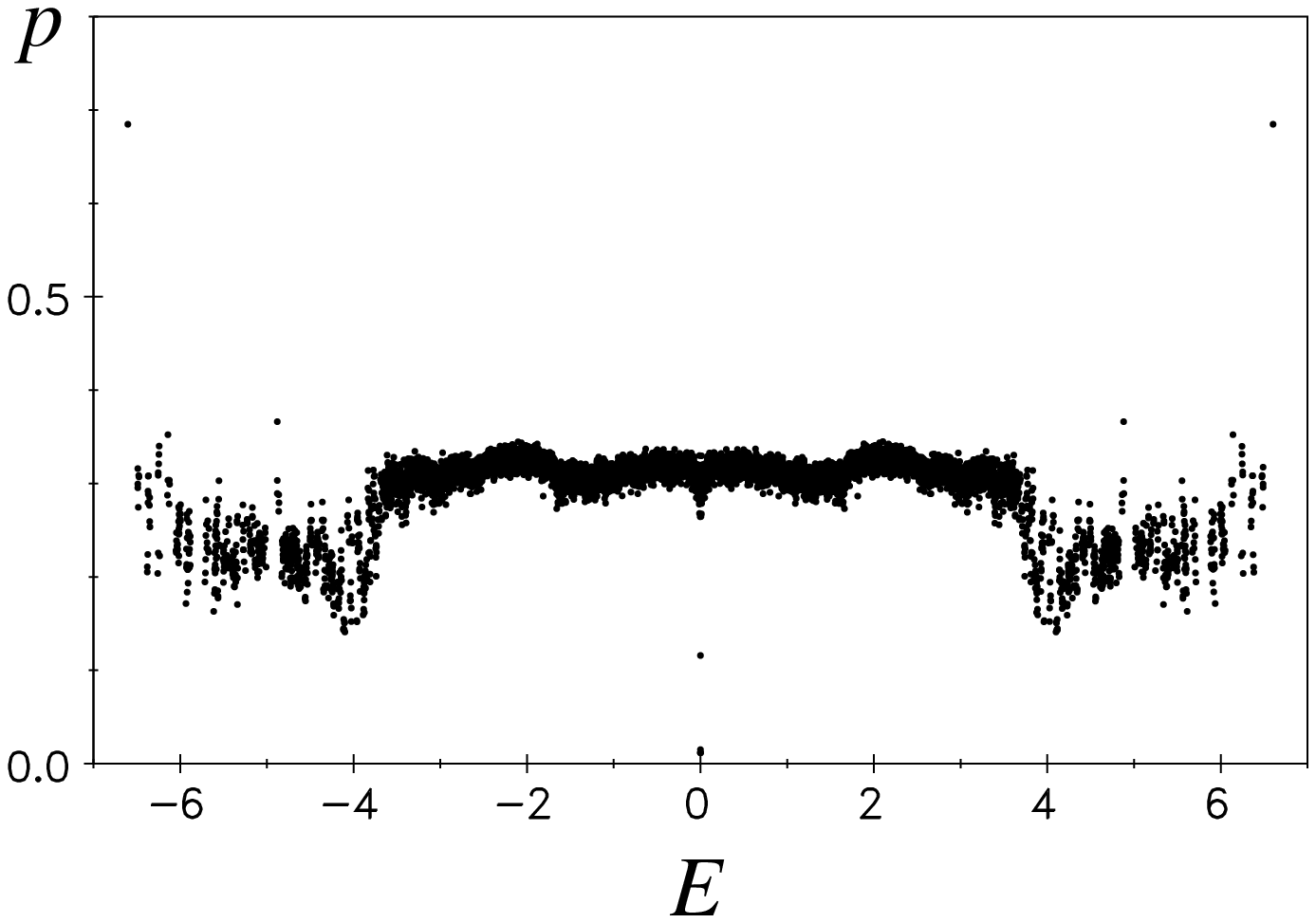}}
\vchcaption{Participation ratios $p=P/N$ for eigenstates of periodic
approximants of the rhombic Penrose tiling (top) and the
Ammann-Kramer-Neri tiling (bottom).}
\label{fig:pr}
\end{vchfigure}

In general, it is expected that generalised eigenstates of
tight-binding Hamiltonians on aperiodic tilings of the plane, and
probably also in three dimensions, are also critical, i.e., neither
exponentially localised nor extended.\index{states!critical}
\index{states!extended}\index{states!localised} However, no
mathematically rigorous results exist as yet, so conclusions are based
on numerical observations.  A simple quantity that characterises the
degree of localisation of a generalised eigenstate with amplitudes
$\psi_{j}$ at vertex $j$ is the participation number $P$ defined
by\index{participation number}
\begin{equation}
P^{-1} = \sum_{j=1}^{N} |\psi_{j}|^4,
\end{equation}
where $N$ denotes the total number of vertices in our finite
approximant. The participation number tells us how many vertices carry
a significant part of the probability measure given by
$|\psi_{j}|^2$. The ratio $p=P/N$, the participation ratio, thus
contains a crude information about the degree of localisation of the
wave packet described by the state $\psi_{j}$. Numerical results for
large periodic approximants of the Penrose and the Ammann-Kramer-Neri
tiling are shown in Fig.~\ref{fig:pr}. They indicate that, for the
system size under consideration, the participation ratio lies around
$0.3$, and there appears to be a tendency towards a stronger
localisation in the ``band'' centre for the Penrose tiling, while in
the icosahedral case states near the ``band'' edges appear to be more
localised.

\begin{vchfigure}[tb]
\centerline{\includegraphics[width=0.39\textwidth]{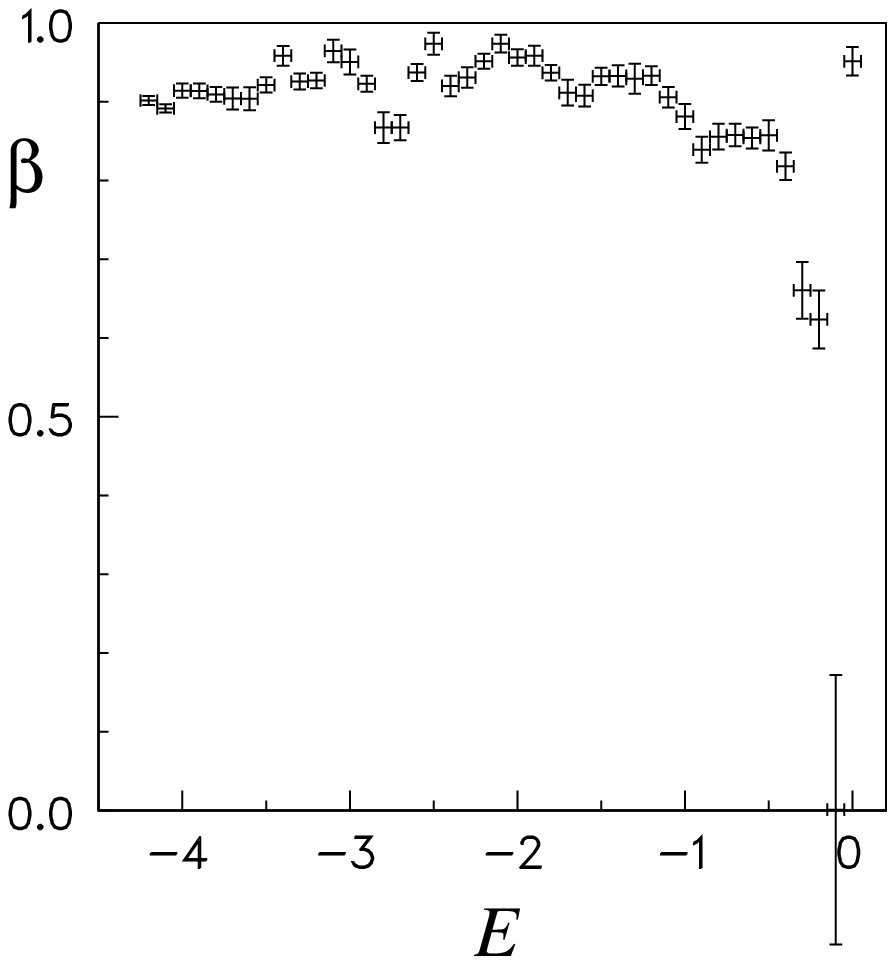}
\hspace*{0.05\textwidth}
\includegraphics[width=0.38\textwidth]{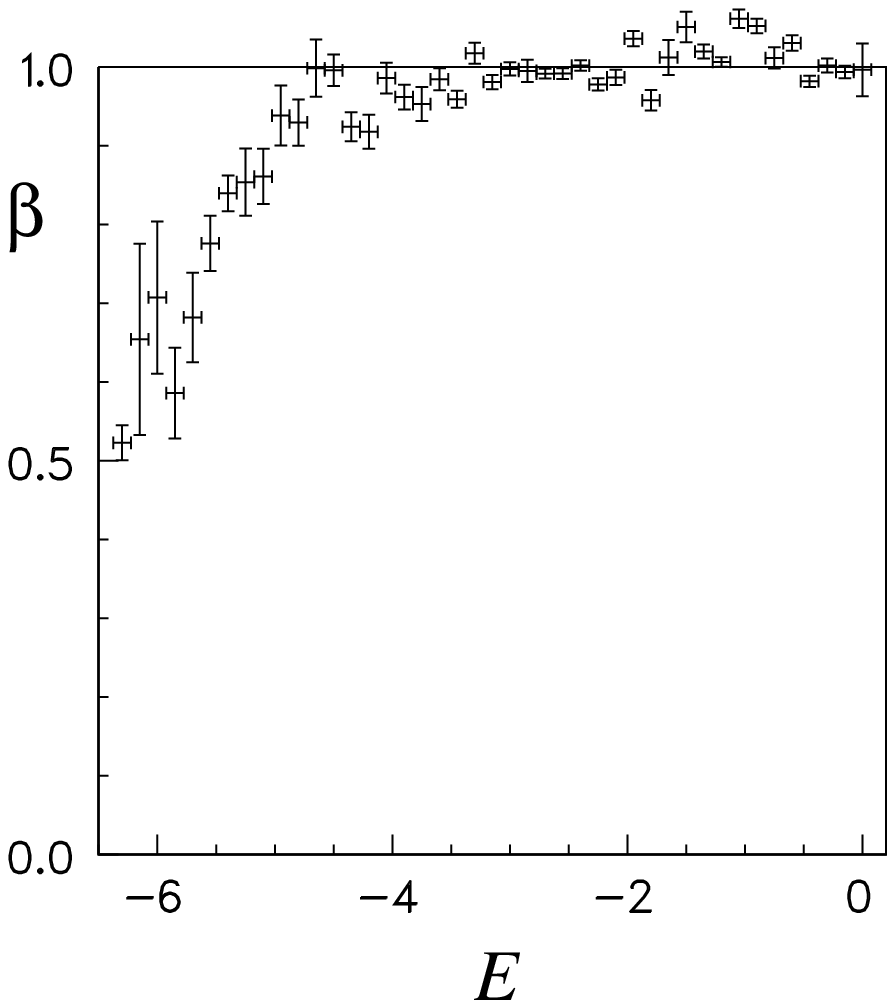}}
\vchcaption{Estimated scaling exponents $\beta$ of the participation 
number for eigenstates of the Penrose tiling (left) and of the
Ammann-Kramer-Neri tiling (right).}
\label{fig:pb}
\end{vchfigure}

However, what we actually need to know in order to obtain information
about localisation properties in the infinite tiling is the scaling
behaviour of the participation number with the size of the
approximant. For an extended state, we expect $P$ to grow linearly
with $N$, whereas for a localised state $P$ will eventually be
constant. For our critical states, which sometimes are also referred
to as algebraically localised states, we expect a scaling behaviour
$P\sim N^{\beta}$ with some exponent $0\le \beta\le 1$. Note that
$\beta=0$ for exponentially localised states, and $\beta=1$ for
extended states; so any value $0<\beta<1$ points towards the presence
of critical states.  The converse is not true --- if $\beta=0$ or
$\beta=1$ we cannot immediately deduce that the state is exponentially
localised or extended, respectively, as we only consider one moment of
the distribution; for instance, a sufficiently rapid sub-exponential
decay may still give $\beta=0$. Numerical results, again for the
Penrose and the Ammann-Kramer-Neri tiling, are displayed in
Fig.~\ref{fig:pb}. The values of the exponent $\beta$ for the Penrose
tiling are about $0.9$, clearly below one, for most energies except
near the centre of the ``band'' where they are smaller. This is
consistent with a preponderance of multifractal eigenstates which are
neither extended nor exponentially localised. In the three-dimensional
case, the result is less obvious. Whereas states near the ``band''
edge clearly show $\beta<1$, the majority of states yields values for
the exponent $\beta$ which are consistent with extended states. In
these cases, it may be that the system size is simply insufficient to
resolve values of $\beta$ close to, but smaller than one. However, we
also cannot exclude the possibility that $\beta=1$ for a large part of
the spectrum.  Still, as mentioned previously, this does not
automatically imply that the eigenstates are extended.

\begin{vchfigure}[tb]
\centerline{\parbox[b]{0.64\textwidth}{%
\includegraphics[width=0.64\textwidth]{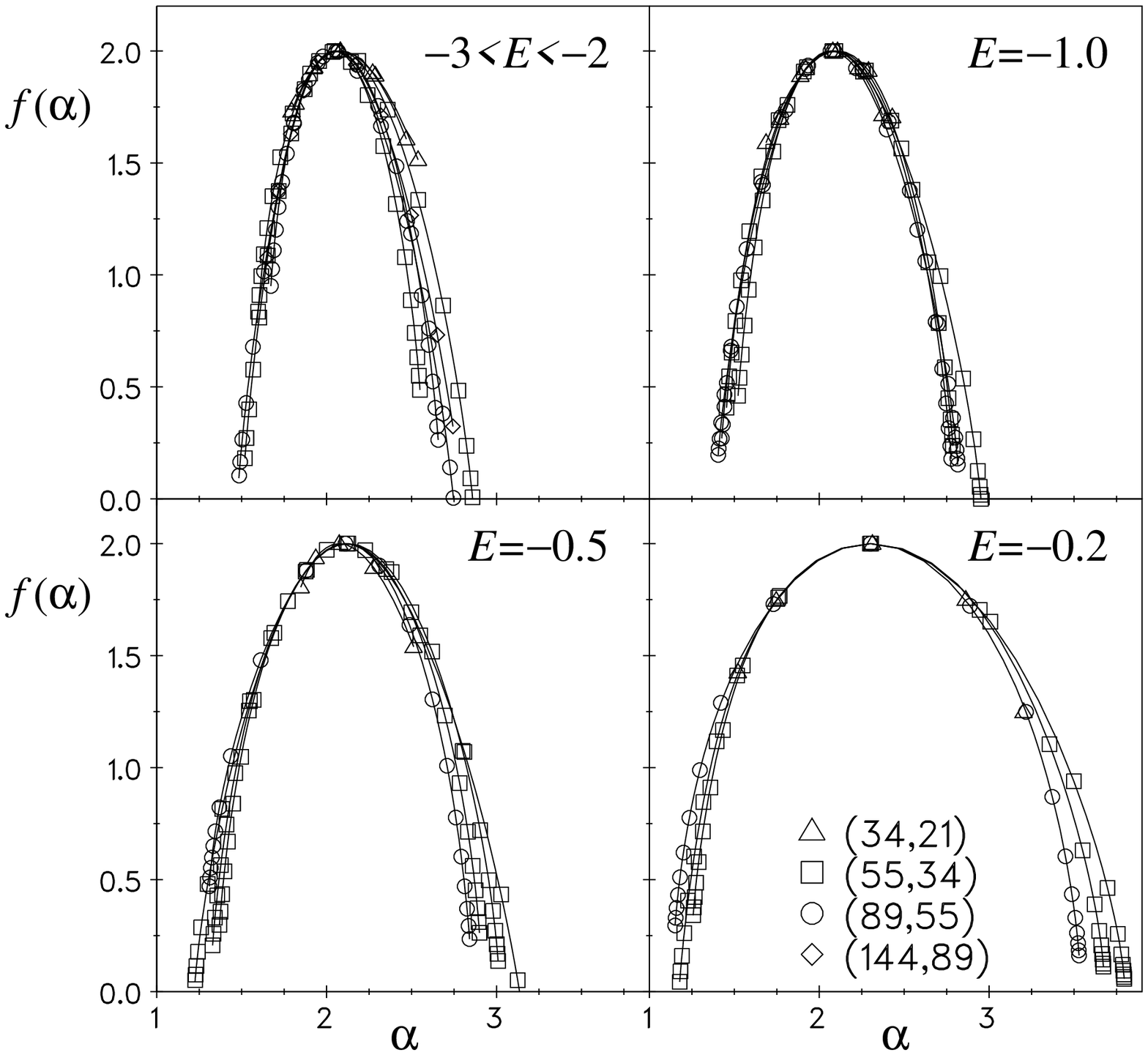}}%
\parbox[b]{0.36\textwidth}{
\vchcaption{Singularity spectra $f(\alpha)$ for various
eigenstates of periodic approximants of the Penrose tiling.
The approximants are labelled by two integers whose ratio is a
rational approximation of the golden mean $\tau=(1+\sqrt{5})/2$.\bigskip}
\label{fig:pfa}}}
\end{vchfigure}

\begin{vchfigure}[tb]
\centerline{\parbox[b]{0.6\textwidth}{%
\includegraphics[width=0.6\textwidth]{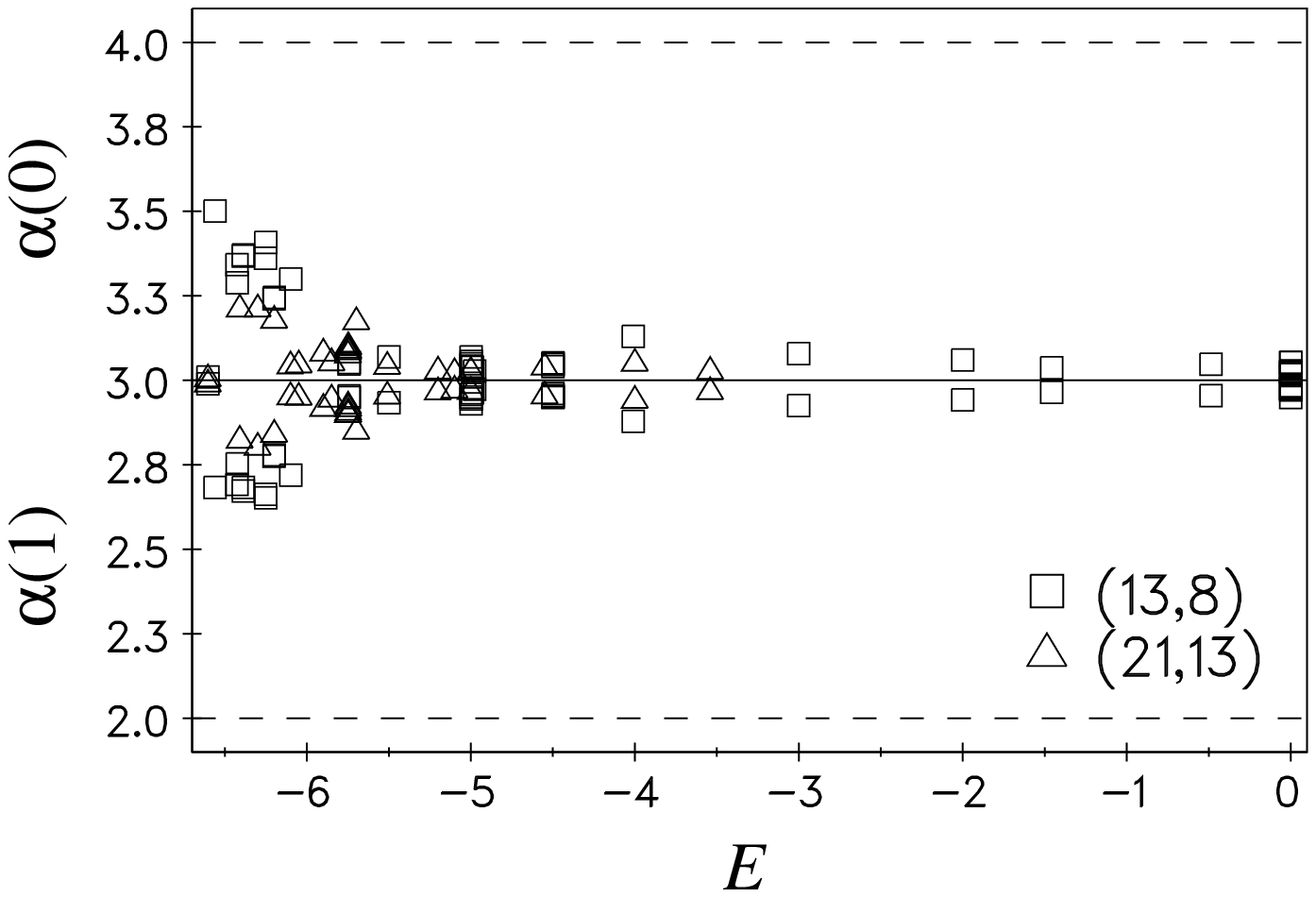}}%
\parbox[b]{0.4\textwidth}{%
\vchcaption{Singularity strengths $\alpha(0)$ and $\alpha(1)$ for
eigenstates of the Ammann-Kramer-Neri tiling. The
approximants are labelled by two integers whose ratio is a
rational approximation of $\tau$.\bigskip}
\label{fig:aknfa}}}
\end{vchfigure}

Another, more powerful approach to characterise ``critical'' states is
given by a multifractal analysis.\index{multifractal
analysis}\index{fractal dimension} In the standard box-counting
approach, the system is divided into boxes of linear size $\delta$. We
consider the measures of the normalised $q$\/th moment
$\mu_k(q,\delta)$ of the probability amplitudes $\mu_k(\delta)$ in the
boxes labelled by $k$.  We obtain the Lipshitz-H\"{o}lder exponent or
singularity strength $\alpha$ of an eigenstate and the corresponding
fractal dimension $f$ by
\begin{equation}
\alpha(q)  =  \lim_{\delta\rightarrow 0}\:
\sum_k\, \frac{\mu_k(q,\delta)\, \ln\mu_k(1,\delta)}{\ln\delta},\quad
f(q)   =  \lim_{\delta\rightarrow 0}\:
\sum_k\, \frac{\mu_k(q,\delta)\, \ln\mu_k(q,\delta)}{\ln\delta},
\end{equation}
yielding the characteristic singularity spectrum $f(\alpha)$ in a
parametric representation. According to \cite{GS95}, the generalized
fractal dimensions $D_{q}^{\psi}$ can be obtained via a Legendre
transformation $D_{q}^{\psi}=\{f[\alpha(q)]-q\alpha(q)\}/(1-q)$.

Some results for the Penrose and Ammann-Kramer-Neri tiling are shown
in Figs.~\ref{fig:pfa} and \ref{fig:aknfa}, respectively
\cite{RS98}. Clearly, eigenstates on the Penrose tiling show
characteristic multifractal behaviour, with singularity spectra
$f(\alpha)$ that are nearly independent of the system size.  Towards
the ``band'' centre, the singularity spectra become wider, indicating
a larger degree of localisation. On the basis of our numerical
analysis, we cannot draw similar conclusions for the icosahedral case,
as the singularity strengths $\alpha(0)=D_{0}^{\psi}$ and
$\alpha(1)=D_{1}^{\psi}$ shown in Fig.~\ref{fig:aknfa} are very close
to $3$, the value for extended states, except near the ``band'' edges.

Summarising the results mentioned so far, we have the situation that
the characterisation of spectra and eigenstates is rather advanced for
one-dimensional aperiodic Schr\"{o}dinger operators, with a number of
mathematically rigorous results available. For planar quasiperiodic
tilings, numerical results strongly favour the conjecture that typical
eigenstates are neither extended nor exponentially localised, but have
multifractal characteristics. In the three-dimensional case, the
situation is less clear. One might expect a rather similar behaviour
as in two dimensions, but numerical results are inconclusive. They do
not rule out that large parts of the spectrum might contain extended
states, although the behaviour of the so-called structural entropy
considered in \cite{RieGS98} hints at a power-law decay. However,
introducing a random energetic disorder (like in the Anderson model)
in the icosahedral tight-binding model leads to a localisation
transition which appears to be very similar to the metal-insulator
transition observed in a simple cubic lattice \cite{RS97}, which might
be regarded as evidence favouring the existence of extended
states. The main problem with all these results is that they may
simply be artifacts of the finite system size which might yet be too
small to resolve the multifractal behaviour.

\subsection{Level-Spacing Distribution}\index{level spacing}

We now return to discuss a rather different property of the energy
spectrum, the statistics of the energy level distribution. The
statistical analysis of energy levels was originally applied to the
complex energy spectra of nuclei, but has since been shown to be
relevant to many complex systems. In the Anderson model of
localisation,\index{Anderson model} the localisation transition is
accompanied by a qualitative change in the normalised distribution
$P^{}_{0}(s)$ of spacings between adjacent energy levels
\cite{HS93a,HS93b,SSSLS}.  For the weakly disordered, metallic phase,
the level-spacing distribution $P^{}_{0}(s)$ is well described by the
corresponding distribution $P^{\rm GOE}_{0}(s)$ of the Gaussian
orthogonal ensemble of random matrix theory \cite{Mehta}, reflecting
the level repulsion or hybridisation of neighbouring extended states.
In the strongly disordered regime, where eigenstates are exponentially
localised, the spacing is described by a Poisson law $P^{\rm
P}_{0}(s)=\exp(-s)$, because the localised states can be arbitrarily
close in energy. These results are universal in the sense that they do
not depend on details of the model, but are a general feature of a
large class of systems sharing a few general properties.\index{random
matrices}

\begin{vchfigure}[tb]
\centerline{\parbox[b]{0.5\textwidth}{%
\includegraphics[width=0.5\textwidth]{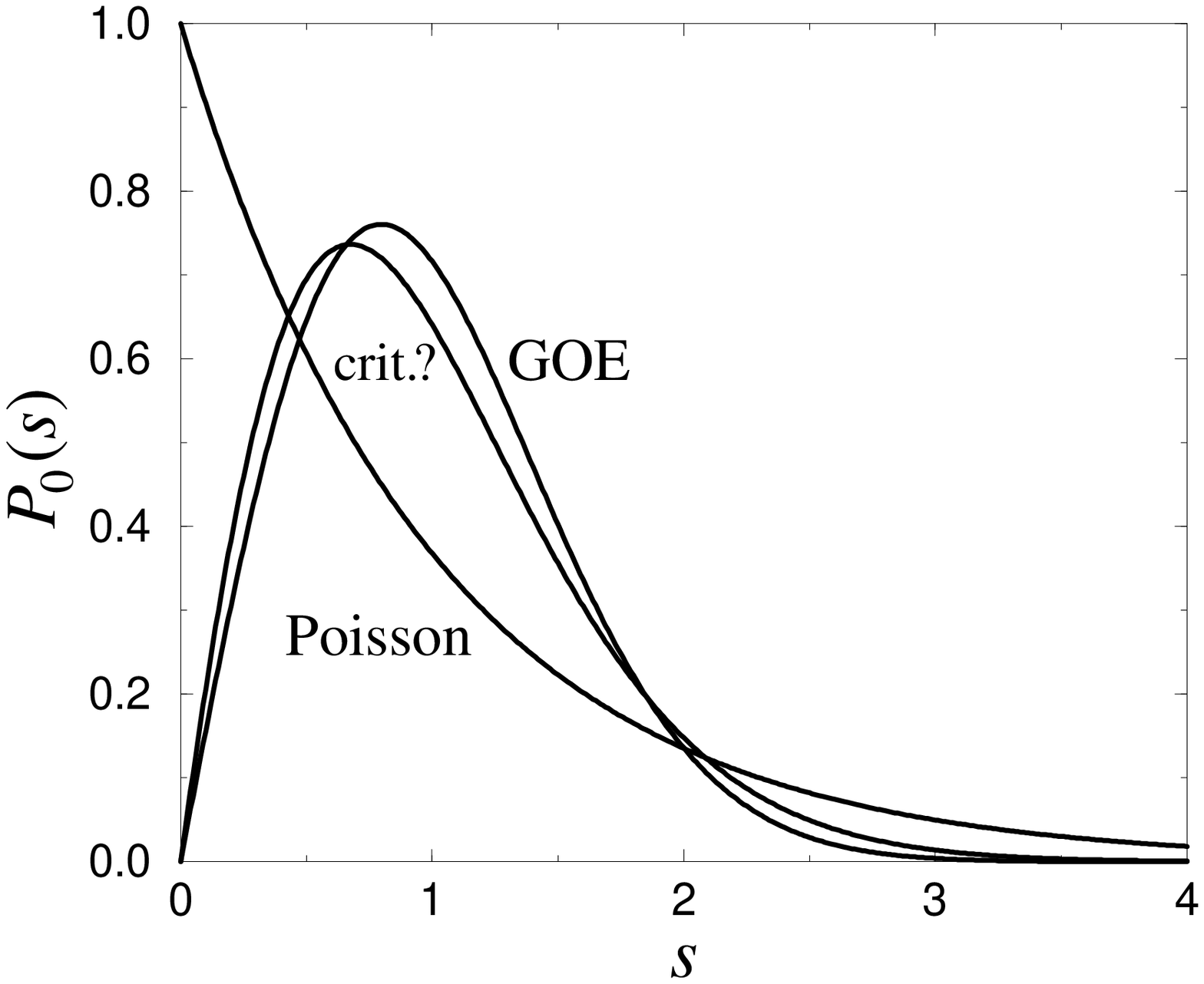}}%
\parbox[b]{0.5\textwidth}{\vchcaption{The level-spacing distribution
$P^{\rm GOE}_{0}(s)$ of the Gaussian orthogonal ensemble of random
matrix theory, the Poisson law $P^{\rm P}_{0}(s)=\exp(-s)$ and a
suggested ``critical'' distribution \cite{ZK97}\bigskip.}
\label{fig:lsd}}}
\end{vchfigure}

At the metal-insulator transition,\index{metal-insulator transition}
the level-spacing distribution has been shown to follow yet another
behaviour which is attributed to the existence of a ``critical
ensemble'' \cite{ZK97}. In contrast to the other two cases, this
``critical'' level-spacing distribution appears to be non-universal. A
sketch of the three level-spacing distributions is given in
Fig.~\ref{fig:lsd}, where the critical curve corresponds to the
function discussed in \cite {ZK97}.

As discussed above, eigenstates in planar quasiperiodic tight-binding
models appear to be generically
multifractal,\index{states!multifractal} and thus similar to the
eigenstates found at the metal-insulator transition of the Anderson
model of localisation.\index{Anderson model} In the latter model it is
extremely difficult to do level-spacing analysis at
criticality,\index{level spacing} because only states near the
mobility edge may be taken into account. In planar quasiperiodic
tight-binding models, however, we have a large reservoir of
multifractal states, and thus we might expect to find some
``critical'' statistics intermediate between the Poisson behaviour and
the universal random matrix distribution.\index{random matrices}

Indeed, early investigations found significant deviations from the
random matrix behaviour \cite{BS91,PJ95}.  However, the system
considered in these papers is a standard periodic approximant of the
Ammann-Beenker tiling \cite{AGS92,BGM02,BGM02e,GS02} which is a
singular patch with an exact reflection symmetry along a diagonal, but
with the property that the fourfold rotational symmetry is broken only
``weakly''. In Fig.~\ref{fig:ab}, such an approximant is shown,
overlaid with a copy rotated by $90$ degrees. This shows that
mismatches occur only along ``worms''.  Although this is no exact
symmetry, and hence the energy spectrum does not split into
independent sectors, it may influence the level-spacing distribution
and thus lead to non-generic results \cite{ZGRS98}.\index{level
spacing}

\begin{vchfigure}[tb]
\centerline{\includegraphics[width=0.69\textwidth]{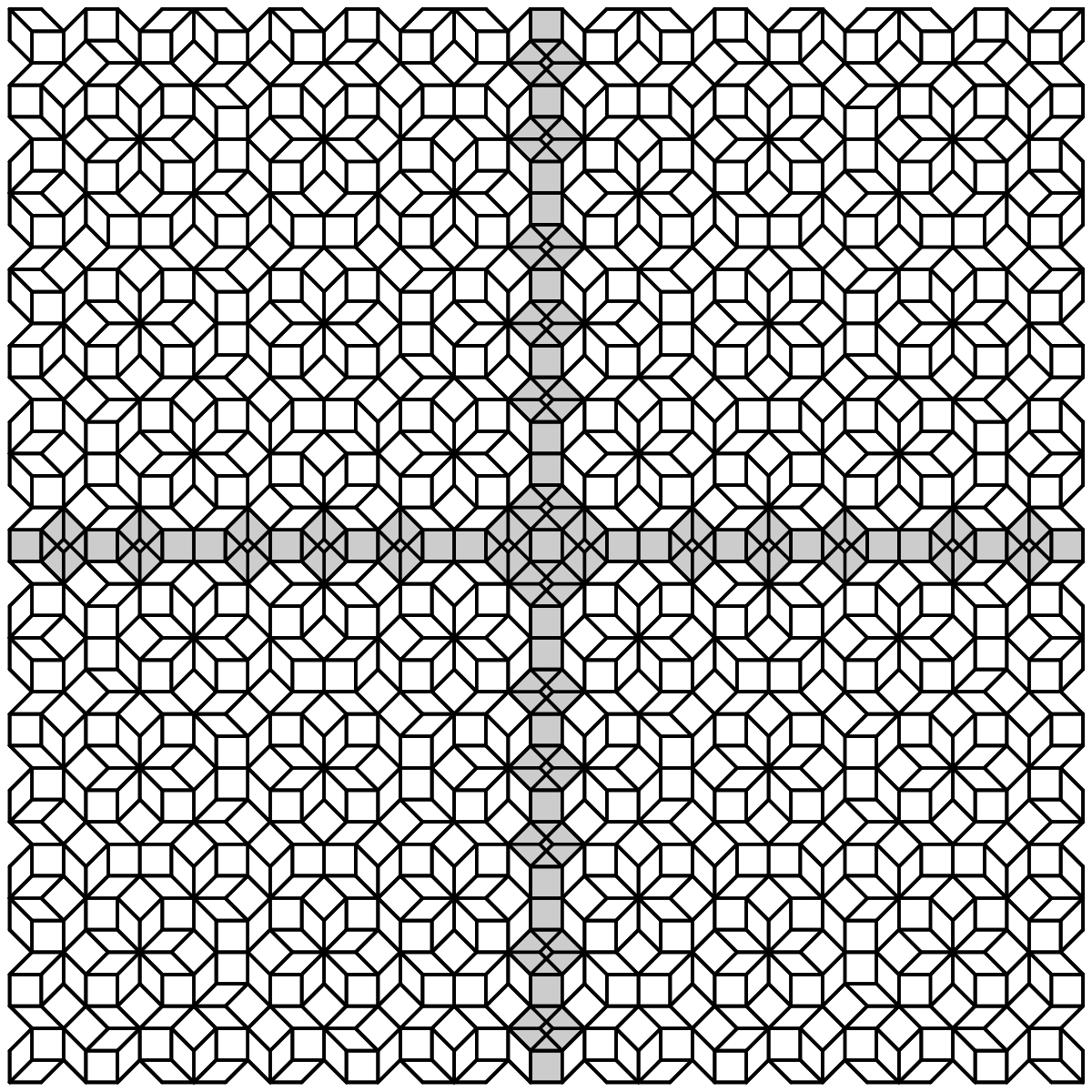}}
\vchcaption{Unit cell of a periodic approximant of the
Ammann-Beenker tiling, overlaid by a copy of itself rotated
by $90$ degrees. Mismatches only occur in the grey shaded ``worms''
\cite{GS02}.\index{Ammann-Beenker tiling}}
\label{fig:ab}
\end{vchfigure}

A careful re-investigation of the energy spectrum\index{spectrum} for
the tight-binding Hamiltonian on the Ammann-Beenker tiling, where the
on-site energies are chosen to be zero and hopping elements are one
along the edges of the tiling, shows that for generic patches the
level-spacing distribution is in fact very well described by the
random matrix distribution \cite{ZGRS98,SGRZ99a,SGRZ99b,GRSZ00}. In
Fig.~\ref{fig:pn}, the numerical results obtained by diagonalising the
tight-binding Hamiltonian for an eightfold symmetric patch of the
Ammann-Beenker tiling with $157\,369$ vertices are shown. In this
case, we need to consider a single irreducible sector corresponding to
one of the ten irreducible representations of the exact $D_{8}$
symmetry of the patch. Here, we not only considered the normalised
spacing distributions $P^{}_{0}(s)$, in terms of the mean level
spacing $s$, for adjacent energy levels, but also the corresponding
spacing distributions $P^{}_{n}(s)$ of pairs of energy levels such
that the energy interval between the two states contains $n$ further
levels. The numerical results for $n=0,1,2,3$ are shown in
Fig.~\ref{fig:pn} and compared with the universal spacing
distributions $P^{\rm GOE}_{n}(s)$ of the Gaussian orthogonal random
matrix ensemble \cite{Mehta,GRSZ00}. The agreement is extremely good,
considering the finite size of the patch. This was already noticed in
\cite{SGRZ99a} where it was shown that the exact random matrix
distribution $P^{\rm GOE}_{0}(s)$ fits the numerical distribution
better than Wigner's surmise $P^{\rm W}_{0}(s)=\pi s\exp(-\pi
s^2/4)/2$, which is a good approximation of $P^{\rm GOE}_{0}(s)$. This
result has also been verified for different patches and boundary
conditions \cite{ZGRS98,SGRZ99a,SGRZ99b,GRSZ00}, and other planar
quasiperiodic systems \cite{ZG99}. Similar results were also found in
topologically disordered systems \cite{GRS98}.

There is, however, a non-trivial step involved in extracting the
spacing distributions shown in Fig.~\ref{fig:pn} from the raw
eigenenergies of the finite system. As Fig.~\ref{fig:dosidos} shows,
the DOS varies considerably over the spectrum, and we have to correct
for this variation if we wish to compare with the universal random
matrix distributions. This process, known as
``unfolding'',\index{spectrum!unfolding} is rather tricky in our case
because it requires a clear distinction between different scales. On
the one hand, we need to average out the fluctuations in the DOS, so
we average on a scale that is given by the fluctuations, see
Fig.~\ref{fig:dosidos}. On the other hand, we are looking at the
spacing distribution of energy levels, so averaging has to be done on
a scale that is large compared to the mean level spacing. It is not
easy to fulfill these two requirements for planar quasiperiodic
tight-binding models, because the two scales do not seem to be well
separated, at least for relatively small systems, compare also
\cite{J02} for an approach based on the inflation symmetry of the
tiling. Here, we used a simple unfolding procedure by approximating
the IDOS by a smooth spline function
\cite{ZGRS98,SGRZ99a,SGRZ99b,GRSZ00}.

\begin{vchfigure}[tb]
\centerline{\parbox[b]{0.55\textwidth}{%
\includegraphics[width=0.55\textwidth]{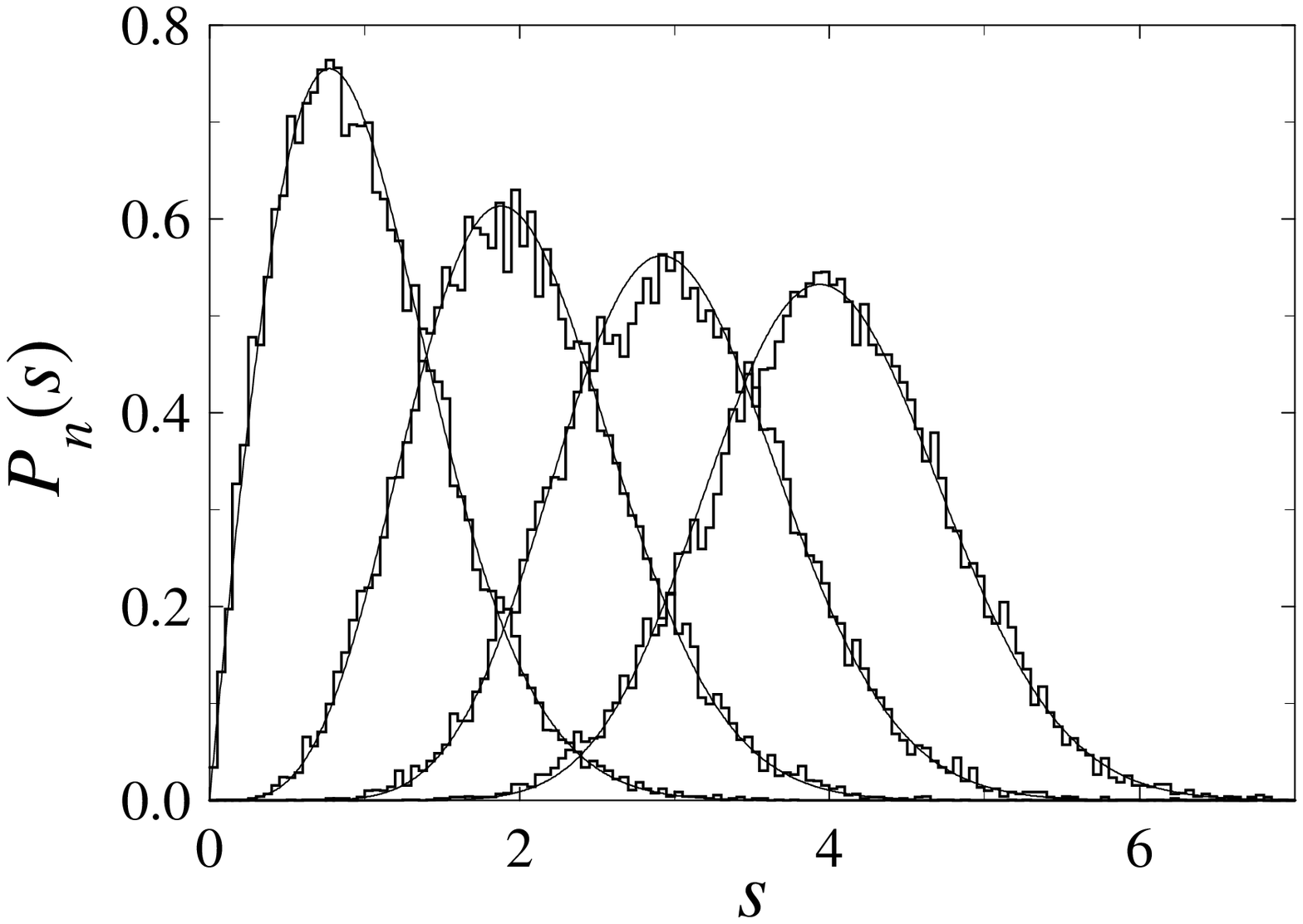}}%
\parbox[b]{0.45\textwidth}{%
\vchcaption{The histograms show the level-spacing distributions
$P^{}_{n}(s)$, $n=0,1,2,3$ (from left to right), as obtained from
the unfolded numerical spectrum of the tight-binding Hamiltonian
on an eightfold symmetric patch of the Ammann-Beenker tiling 
containing $N=157\,369$ vertices. The smooth curves are the 
corresponding level-spacing distributions $P^{\rm GOE}_{n}(s)$.
\bigskip}\label{fig:pn}}} 
\end{vchfigure}

\begin{vchfigure}[b]
\centerline{\parbox[b]{0.54\textwidth}{%
\includegraphics[width=0.54\textwidth]{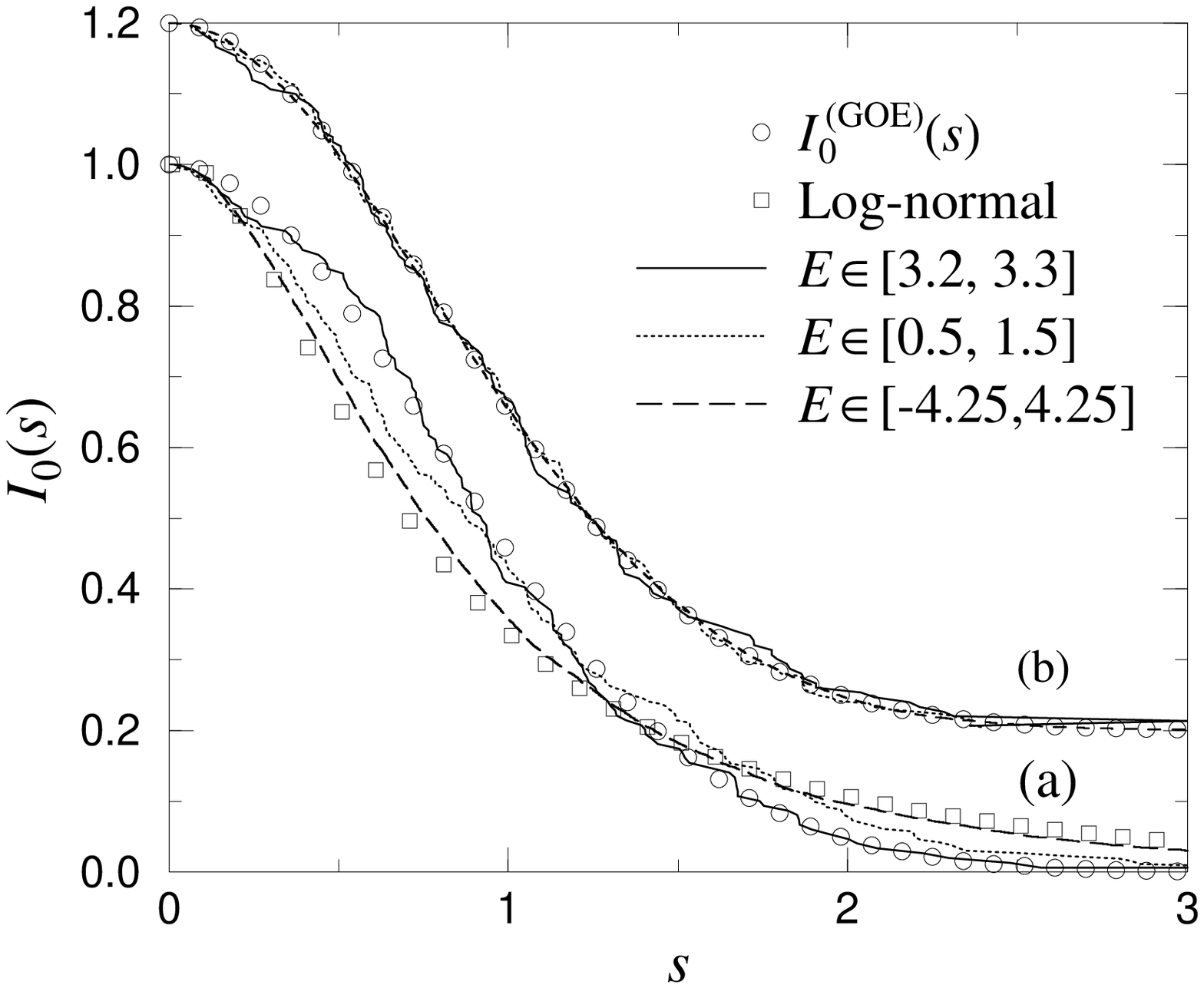}}%
\parbox[b]{0.46\textwidth}{%
\vchcaption{Integrated level-spacing distribution $I^{}_{0}(s)$
obtained (a) without unfolding and (b) with unfolding for various
parts of the spectrum of one sector of the same $D_8$-symmetric patch
as in Fig.~\ref{fig:pn}: whole spectrum (dashed line), $0.5\le E\le
1.5$ (dotted line), and $3.2\le E\le 3.3$ (solid line). Circles and
boxes denote $I^{\rm GOE}_{0}(s)$ and the log-normal distribution,
respectively. The curve for $I^{}_{0}(s)$ in (b) has been shifted by
$0.2$ for clarity.
\bigskip}\label{fig:unfold}}}
\end{vchfigure}

There is one direct way to check that the unfolding procedure does not
introduce artifacts in our results \cite{SGRZ99b}. If we consider the
level-spacing distribution for levels within small energy intervals
such that the DOS is approximately constant on the interval, we do not
need to apply any unfolding procedure and can compare the result
directly with the universal distribution function. However, the price
we pay is that the statistics are much worse, as only the limited
number of eigenenergies in the chosen interval
contributes. Fig.~\ref{fig:unfold} shows the result for three energy
intervals. The numerical results for the integrated level-spacing
distribution $I^{}_{0}(s)=\int_{s}^{\infty}P^{}_{0}(t)\,\mathrm{d}t$,
with and without unfolding, are compared with the log-normal
distribution favoured in \cite{BS91,PJ95} and with the integrated
level-spacing distribution $I^{\rm GOE}_{0}(s)$ of the Gaussian
orthogonal random matrix ensemble.

Comparing the energy intervals considered in Fig.~\ref{fig:unfold}
with the DOS displayed in Fig.~\ref{fig:dosidos}, it is apparent that
spacing distributions for an approximately constant DOS (for instance,
for $3.2\le E\le 3.3$) exhibit random matrix behaviour even without
unfolding. However, for energy ranges with fluctuating DOS, the
integrated level-spacing distribution $I^{}_{0}(s)$, without
unfolding, deviates from $I^{\rm GOE}_{0}(s)$. In the interval $0.5\le
E\le 1.5$ with large fluctuations, the level-spacing distribution of
the raw eigenenergies is neither well described by a log-normal
distribution nor by $I^{\rm GOE}_{0}(s)$. For the entire spectrum, the
distribution of the raw spacings is actually close to a log-normal
distribution, but it is our interpretation that this is due to the
abundance of large spacings due to the fluctuations in the
DOS. Clearly, the unfolded distribution functions for the three energy
intervals agree well with each other and with the spacing distribution
of the Gaussian random matrix ensemble.

In conclusion, the spectra of planar quasiperiodic tight-binding
models of the type considered here appear to obey random matrix
statistics.\index{random matrices} Compared to the results for the
Anderson model,\index{Anderson model} this means that, at least in
this respect, they behave like weakly disordered systems rather than
like systems at a metal-insulator transition,\index{mertal-insulator
transition} in spite of their multifractal
eigenstates.\index{states!multifractal}

\subsection{Multifractal Eigenstates on the Penrose Tiling}
\index{states!multifractal}

We now come back to the eigenstates of planar quasiperiodic
tight-binding models.\index{tight-binding model} As mentioned above,
numerical results indicate that typical states of hopping models on
planar quasi\-periodic tilings are multifractal, with certainly some
exceptions such as single extended states at the ``band'' edges and
confined states in the ``band'' centre. For the rhombic Penrose
tiling, the confined states can be explicitly constructed, see
\cite{RS95} and references therein, and they make up a sizeable
fraction of the spectrum. This is also the case for the Ammann-Beenker
tiling as can be seen from the pronounced peak in the ``band'' centre
in Fig.~\ref{fig:dosidos}.

\begin{vchfigure}[b]
\centerline{\parbox[b]{0.5\textwidth}{%
\includegraphics[width=0.5\textwidth]{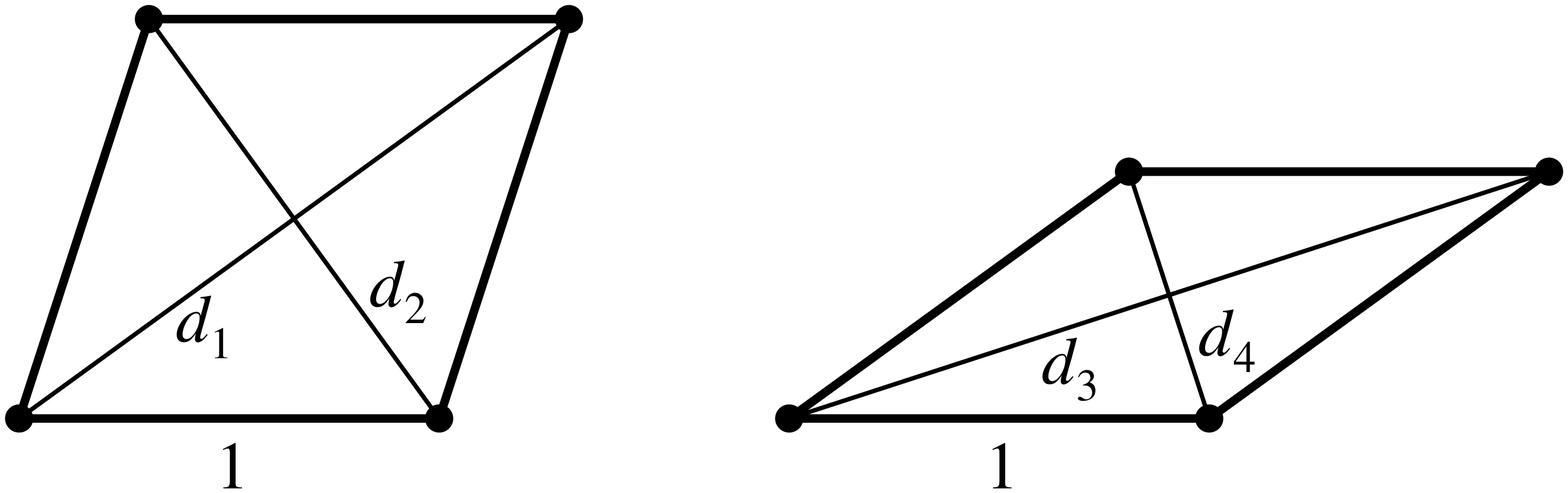}}%
\parbox[b]{0.5\textwidth}{%
\vchcaption{The hopping elements for our tight-binding model on
the Penrose tiling are chosen as $t_{jk}=1$ along the edges of the
tiling, and as $t_{jk}=d_{1},d_{2},d_{3},d_{4}$ along the four
different diagonals of tiles.} \label{fig:rhombs}}}
\end{vchfigure}

However, it is also possible to construct some multifractal
eigenstates on the Penrose tiling explicitly. This can be achieved by
following an ingenious idea of Sutherland \cite{S86} to exploit the
matching rules of the tiling to derive a non-trivial ansatz for a
multifractal wave function.\index{states!multifractal} Based on this
idea, non-normalisable eigenstates of the centre model, where
electrons may hop between neighbouring tiles rather than neighbouring
vertices, were derived, and their multifractal properties were
characterised \cite{TFA88}.

\begin{vchfigure}[tb]
\centerline{\includegraphics[width=0.82\textwidth]{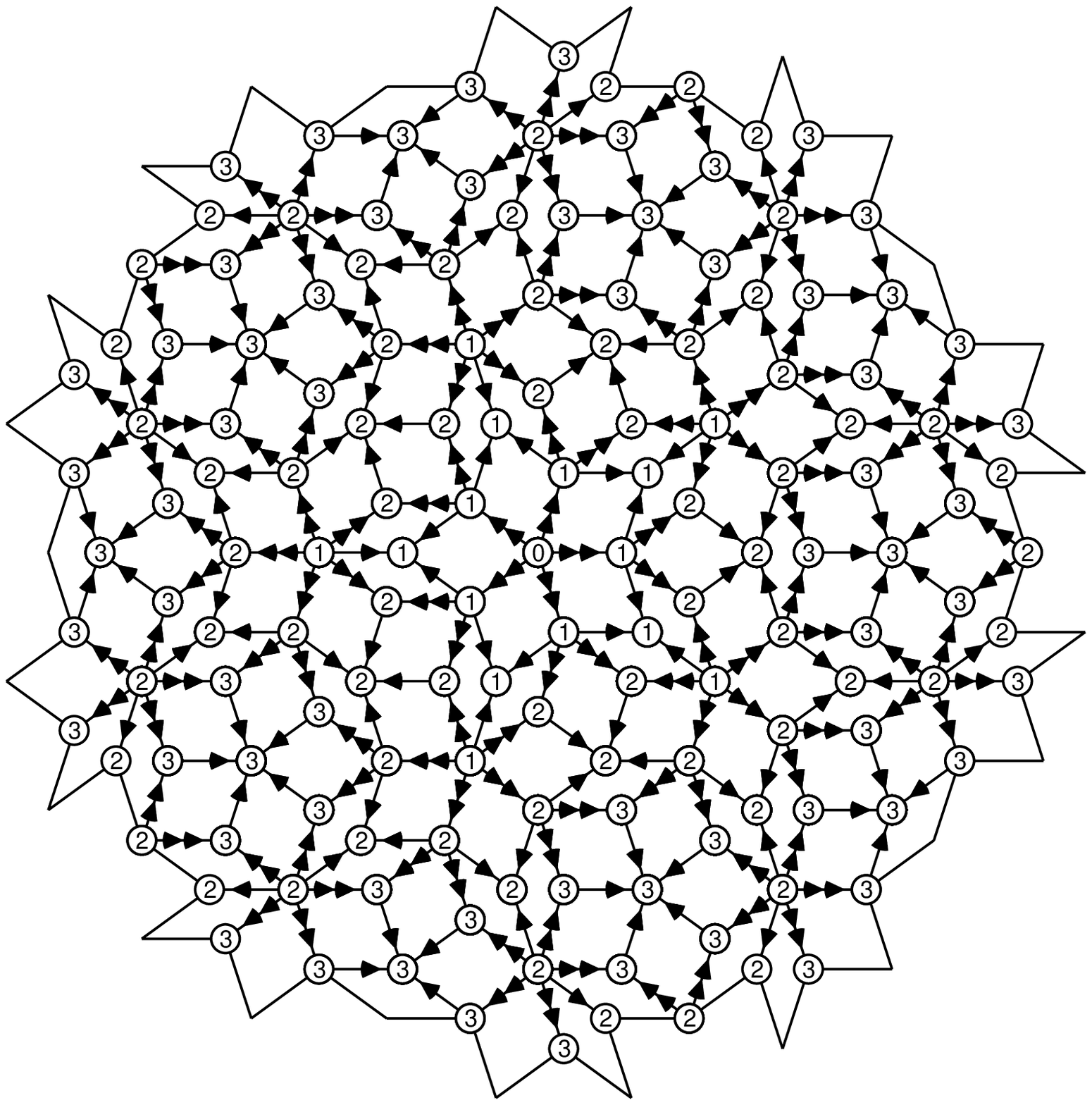}}
\vchcaption{Arrow decoration for vertices $k$ on a patch of the
rhombic Penrose tiling encoding the matching rules and the values of
the height function $m(k)$ with respect to the centre
vertex.\index{Penrose tiling}\index{matching rules}}
\label{fig:pot}
\end{vchfigure}

Here, we apply the same idea to the vertex model on the Penrose
tiling, i.e., electrons may hop from one vertex of the Penrose tiling
to its neighbouring vertices. It turns out that we have to generalise
our model slightly in order to find non-trivial solutions
\cite{RepGS98}. Therefore, we include hopping along the diagonals of
the rhombic tiles, with hopping parameters in the Hamiltonian
(\ref{eq:h}) chosen as $t_{jk}=1$ between two vertices $j$ and $k$
connected by an edge of the tiling, $t_{jk}=d_{1},d_{2},d_{3},d_{4}$
between vertices $j$ and $k$ on the four different diagonals of the
two rhombs, see Fig.~\ref{fig:rhombs}, and $t_{jk}=0$ otherwise, and
on-site energies $\varepsilon_{k}=0$.

As mentioned, the ansatz for the eigenstates of our Hamiltonian on the
infinite tiling involves the matching rules of the Penrose
tiling.\index{matching rules} These are usually encoded in a
decoration of the tiling with two types of arrows, single and double
arrows, located on the edges of the tiles, see Fig.~\ref{fig:pot} for
an example.  The matching rules require that two tiles can only share
an edge if the corresponding arrow decorations match, both in type and
direction.  Given such a decoration of the infinite tiling, we define
a ``potential'' or ``height function'' as follows. First, we choose an
arbitrary vertex $k_{0}$ on the tiling and define the height of this
vertex as $m(k_{0})=0$. Then, for any vertex $k$, we consider any path
consisting of a sequence of edges that connects $k_{0}$ to $k$.  The
height $m(k)$ is given by counting the number of double arrows
encountered along the path, where double arrows pointing along the
direction from $k_{0}$ to $k$ count as $1$, whereas arrows counting in
the opposite direction count as $-1$. This is well defined because for
any closed path this number is zero, as can easily be verified for the
two basic tiles. The height at any vertex is thus an integer
number. Its actual value depends on our choice of $k_{0}$, but any
other choice $k_{0}^{\prime}$, defines a height function $m^{\prime}$
that differs from the potential or height function $m$ defined by
$k_{0}$ only by a constant $m(k_{0}^{\prime})$.  Note that the word
``potential'' is used in a different context here, there are no
on-site potentials in our Hamiltonian. The height function $m(k)$ is
going to appear in the ansatz for the eigenfunction.

For the Hamiltonian on the infinite tilings, we need to solve the
infinite set of equations
\begin{equation}
\sum_{k} t_{jk}\,\psi_{k} = E\psi_{j}, \label{eq:tb}
\end{equation}
where $t_{jk}=t_{kj}\in\{0, 1, d_{1},d_{2},d_{3},d_{4}\}$ as described
above. Our ansatz for the amplitudes $\psi_{j}$ of the wave function
at vertex $j$ involves the height function $m(j)$ and information
about the local neighbourhood of the vertex $j$ \cite{RepGS98}. In the
simplest case, we choose
\begin{equation}
\psi_{j}=A_{\nu(j)}\,\gamma^{m(j)},
\label{eq:ansatz}
\end{equation}
where $A_{\nu(j)}$ are eight constants corresponding to the eight
different vertex stars $\nu(j)$ in the decorated Penrose tiling, and
$\gamma$ is another parameter. Altogether we have nine free parameters
in the ansatz, as well as the four hopping parameters
$d_{1},d_{2},d_{3},d_{4}$ and the energy $E$. Inserting the ansatz
(\ref{eq:ansatz}) into our infinite set (\ref{eq:tb}) of equations
reduces the number of independent equations for our parameters to a
finite number, namely 31 equations. Essentially, the equations can be
labelled by the different vertex types in the Penrose tiling taking
into account the next-nearest neighbours, as for each such patch
around a vertex $j$ all vertex types and values of the height function
for the vertices $k$ that enter Eqs.~(\ref{eq:tb}) are determined.

Even though the number of equations is still more than twice as large
as the number of parameters, there exist indeed solutions
\cite{RepGS98}.  It turns out that two different amplitudes
$A_{\nu(j)}$ suffice in the ansatz (\ref{eq:ansatz}), and the wave
functions for our solutions do not depend on the vertex type $\nu(j)$,
but only on the translation class $t(j)$. The vertices of the Penrose
tiling fall into four such translation classes, corresponding to the
four disjoint parts of the window in the description as a
four-component model set \cite{BM00,SSH02,RepGS99}. It follows from
Eqs.~(\ref{eq:tb}) that the coefficients $A_{\nu(j)}$ have to coincide
for vertices of translation classes $t(j)\in\{1,4\}$, which correspond
to the small pentagon as a window, and for vertices of translation
classes $t(j)\in\{2,3\}$, which correspond to the large pentagon as a
window.

There are three sets of parameters that solve the eigenvalue equations
of (\ref{eq:tb}). It is convenient to introduce the notation
$c_{\pm}=1/(\gamma\pm\gamma^{-1})$ as this combination appears
repeatedly in the expressions for the parameters.  Explicitly, the
three solutions look as follows.

\begin{vchfigure}[tb]
\centerline{\includegraphics[width=0.82\textwidth]{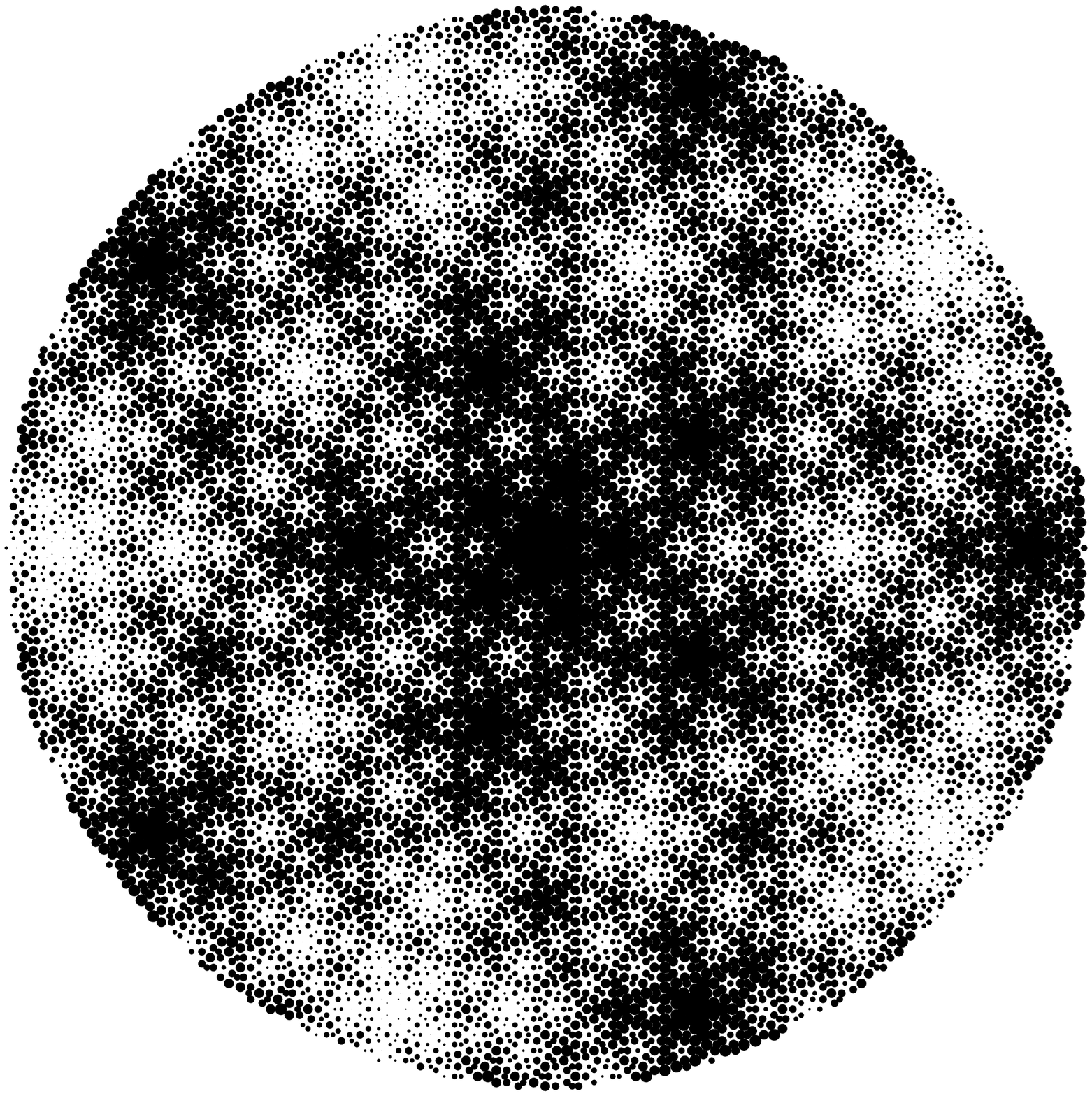}}
\vchcaption{Sketch of the wave function (\ref{eq:sol1})
with $\gamma=3/5$. The probability density
$|\psi_{j}|^{2}$, encoded in dots of different sizes,
 is shown on a finite patch of ${N}=16\, 757$ vertices.}
\label{fig:wv}
\end{vchfigure}

The first solution has the form
\begin{equation}
\psi_{j}^{(1)} =
\begin{cases}
             (1 - 2\gamma^2)\gamma^{m(j)} & \mbox{for $t(j)\in\{1,4\}$}\\
             \gamma^{m(j)+1}              & \mbox{for $t(j)\in\{2,3\}$}
\end{cases}
\label{eq:sol1}
\end{equation}
with hopping parameters $d_{1}=c_{-}/2$, $d_{2}=-3c_{-}/4+c_{-}^{-1}$,
$d_{3}=-c_{-}/4+c_{-}^{-1}/2$, $d_{4}=c_{-}$, and an energy eigenvalue
$E=-5c_{-}/2$. The other two solutions are given by
\begin{equation}
\psi_{j}^{(2)} =
\begin{cases}
             \gamma^{m(j)}   & \mbox{$t(j)\in\{1,4\}$}\\
             \gamma^{m(j)+1} & \mbox{$t(j)\in\{2,3\}$}
\end{cases},\qquad
\psi_{j}^{(3)} =
\begin{cases}
             \gamma^{m(j)+1} & \mbox{$t(j)\in\{1,4\}$}\\
             \gamma^{m(j)}   & \mbox{$t(j)\in\{2,3\}$}
\end{cases},
\label{eq:sol23}
\end{equation}
with $d_{1}=-c_{+}$, $d_{2}=3c_{+}/2-c_{+}^{-1}/2$,
$d_{3}=-(1+2\gamma^{2})c_{+}/2$, $d_{4}=(1-\gamma^{-2})c_{+}$, and
eigenvalue $E=5c_{+}$ for the solution $\psi_{j}^{(2)}$, and with
$d_{1}=-(1+2\gamma^{2})c_{+}/2$, $d_{2}=c_{+}/4-c_{+}^{-1}/2$,
$d_{3}=-(1+4\gamma^{-2})c_{+}/4$, $d_{4}=-c_{+}$, and eigenvalue
$E=5c_{+}/2$ for $\psi_{j}^{(3)}$, respectively.

These three solutions contain one free parameter, namely $\gamma$. For
any value of $\gamma$, Eqs.~(\ref{eq:sol1}) and (\ref{eq:sol23}) give
solutions of the eigenvalue equations of (\ref{eq:tb}). Note that not
only the hopping parameters $d_{1}$, $d_{2}$, $d_{3}$ and $d_{4}$ are
fixed by $\gamma$, but also the eigenvalue $E$ is determined. In other
words, this means that for a given Hamiltonian, i.e., for a given set
of hopping parameters, our solutions will give at most one
eigenfunction, so we cannot gain any global information on the
spectrum from this result.

An example wave function is shown in Fig.~\ref{fig:wv}. Clearly, the
probability distribution $|\psi_{j}|^{2}$ reflects the topology of the
tiling, because it essentially depends on the height function $m(j)$
only.  The wave functions constructed in this way are not
normalisable; for generic values of $\gamma$ they are neither
exponentially localised nor extended. As expected for the generic wave
functions, they are multifractal.  Generalised dimensions
characterising the eigenstates can be calculated by using the
inflation symmetry of the tiling and investigating the behaviour of
the height function under inflation \cite{RepGS98}.  The results for
the generalised dimensions $D_{q}$ for various values of $\gamma$ are
shown in Fig.~\ref{fig:dq}. The smaller the value of $|\gamma|$, the
faster the wave function decays, giving rise to steeper curves for
$D_{q}$ as a function of $q$.  For $\gamma=1$, the amplitudes
$\psi_{j}$ are independent of the height function $m(j)$, the wave
function is extended, and $D_{q}=2$ is constant.

\begin{vchfigure}[tb]
\centerline{\parbox[b]{0.6\textwidth}{%
\includegraphics[width=0.6\textwidth]{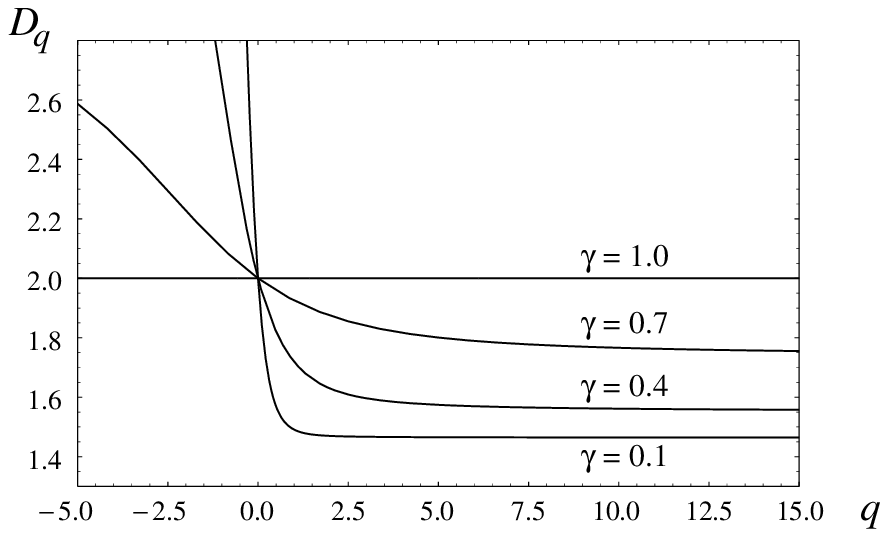}}%
\parbox[b]{0.4\textwidth}{%
\vchcaption{Generalised dimensions $D_{q}$ for the eigenfunction
(\ref{eq:sol1}) constructed by the ansatz (\ref{eq:ansatz}), for
various values of $\gamma$.}\label{fig:dq}\index{fractal dimension}}}
\end{vchfigure}

We note that the ansatz (\ref{eq:ansatz}) can be generalised by taking
into account larger coronae of the vertex $j$. This has been
investigated in \cite{RepGS98}, and it was shown that the eigenvalues
$E$ for the functions constructed here are always infinitely
degenerate. However, numerical investigations indicate that, contrary
to the case of the confined states \cite{RS95}, these states do not
make up a finite fraction of the complete spectrum, so they do not
lead to a peak in the DOS.

In summary, we can construct particular eigenstates for the infinite
Penrose tiling explicitly, following an idea by Sutherland
\cite{S86}. This at least shows that multifractal states exist in the
spectra of such tight-binding Hamiltonians, corroborating the
expectation that these states are generic in planar quasiperiodic
tight-binding models. However, the price to pay is that we only obtain
eigenstates at a particular energy, and thus do not gain any
information about the spectrum. The way in which the matching rules
are used to construct eigenfunctions of a discrete Schr\"{o}dinger
operator is very interesting; it appears plausible that the height
function admits further interpretations as a characteristic feature of
the Penrose tiling.

\subsection{Quantum Diffusion on the Labyrinth}\index{quantum diffusion}

After characterising the spectrum and the eigenstates, we are now
interested in the quantum diffusion in quasiperiodic systems.
Clearly, the diffusion properties of quasicrystals are associated with
the complex eigenstates and energy spectra discussed above
\cite{KPG92}. In what follows, we consider two quantities
characterising the spreading of wave packets \cite{KKKG97}, the
temporal autocorrelation function $C(t)$ and the mean square
displacement $d(t)$. These are defined by
\begin{equation}
C(t) =
\frac{1}{t}\int\limits_{0}^{t}|\psi_{j_{0}}(t^{\prime})|^{2}\,
dt^{\prime},\qquad
 d^2(t) = \sum\limits_{j}|{\bf r}_{j}-
{\bf r}_{j_{0}}|^{2}\,
             |\psi_{j}(t)|^{2},
\label{eq:cd}
\end{equation}
where $\psi_{j}(t)$ is the amplitude of the wavefunction at time $t$
at vertex $j$, which is located at the position ${\bf r}_{j}$ in
space. The function $C(t)$ is the time-averaged probability of a wave
packet staying at the initial site $j_{0}$ at time $t$, whereas $d(t)$
determines the spreading of a wave packet.

Generally, these two functions are characterised by asymptotic power
laws $ C(t)\sim t^{-\delta}$, $d(t)\sim t^{\beta}$ for large time $t$,
where $0<\delta <1$ and $ 0<\beta<1 $ for one-dimensional
quasiperiodic systems \cite{KPG92,KKKG97}.  An intimate relation
between the spectral properties, in particular the fractal dimensions
of the spectrum and the eigenstates, and the exponents $\delta$ and
$\beta$ characterising anomalous diffusion is expected on general
grounds \cite{G89,G93}, and is reasonably well established for
one-dimensional systems. However, only few two-dimensional
quasiperiodic systems have been investigated, notably the
Ammann-Beenker tiling \cite{PSB92} and Fibonacci grids \cite{ZM95},
showing superdiffusive behaviour with an exponent $\beta\geq 1/2$.

\begin{vchfigure}[tb]
\centerline{\includegraphics[width=0.72\textwidth]{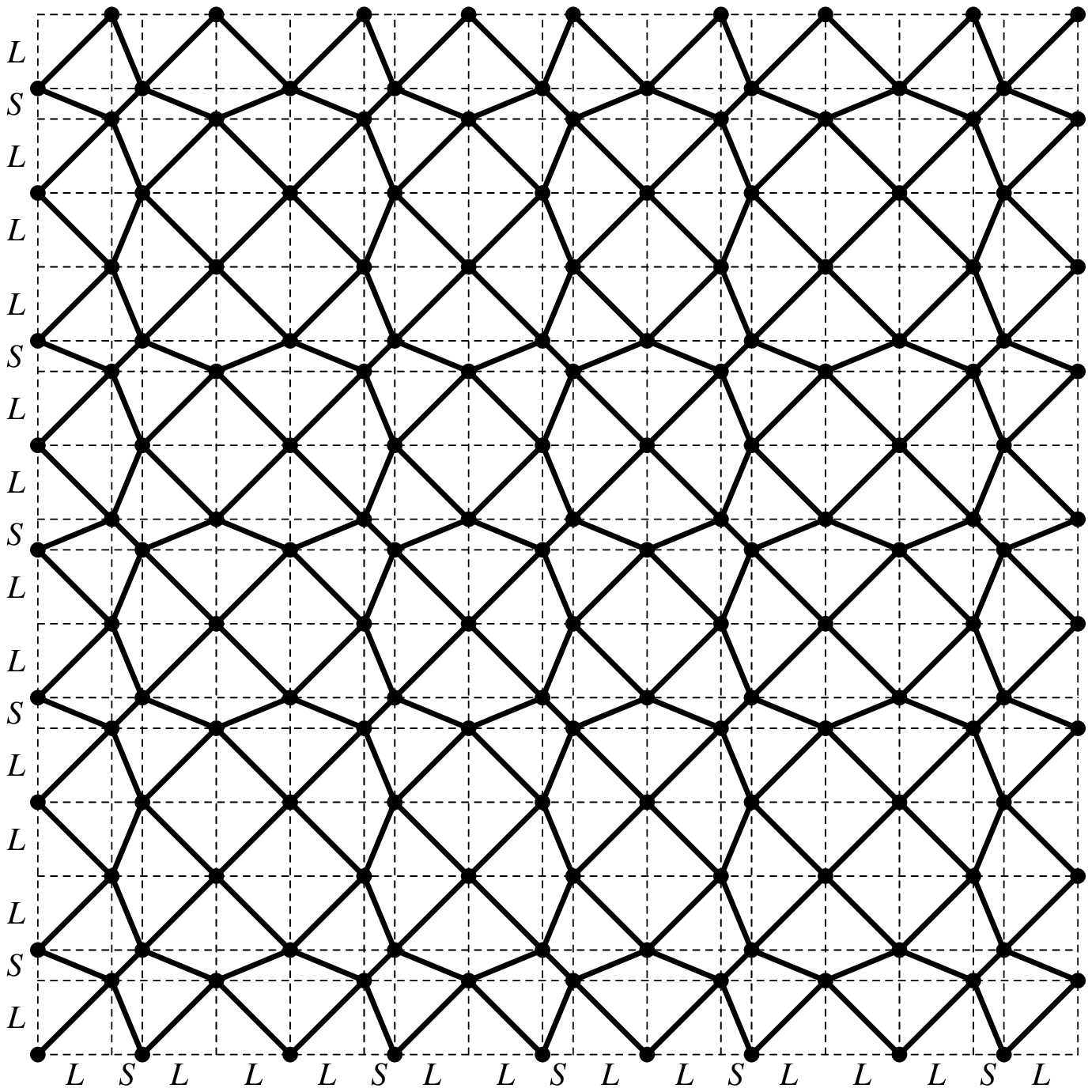}}
\vchcaption{Labyrinth tiling constructed from a product grid
(dashed) of two octonacci chains.\index{labyrinth tiling}} \label{fig:laby}
\end{vchfigure}

Here, we consider a tight-binding model\index{tight-binding model} on
the labyrinth tiling \cite{SMS89}, which is a planar tiling related to
the octagonal tiling \cite{S89}. It is constructed from a rectangular
grid based on the product of two octonacci chains (\ref{eq:subst}), by
connecting vertices that are separated by two steps along the grid,
see Fig.~\ref{fig:laby}. The two letters $L$ and $S$ of the
substitution rule (\ref{eq:subst}) are represented by a long and a
short interval, respectively. Clearly, there are edges of three
different lengths in the labyrinth tiling; the long edges are the
diagonals of large squares formed by two $L$ intervals, the medium
edges diagonals of rectangles formed by one $L$ and one $S$ interval,
and the short edges are the diagonals of small squares formed by two
$S$ intervals.

We define the tight-binding model\index{tight-binding model} on the
labyrinth by the Hamiltonian (\ref{eq:h}) with on-site energies
$\varepsilon_{k}=0$ and non-zero hopping elements $t_{jk}=1$,
$t_{jk}=v$ and $t_{jk}=v^2$ for long, medium and short edges of the
labyrinth tiling, respectively. With this choice, it can be shown that
the eigenfunctions $\psi_{j}$ are essentially products of two
eigenfunctions of the corresponding one-dimensional tight-binding
model on the octonacci chain, with hopping elements $1$ and $v$ for
long and short intervals, respectively \cite{YGRS00, S89}. The
eigenenergies of the labyrinth turn out to be the products of those of
the one-dimensional system.  In other words, the energy spectrum and
the eigenstates of the tight-binding model on the labyrinth tiling can
be obtained from those of the one-dimensional octonacci chain, see
\cite{YGRS00} for details. This allows the numerical treatment of
large systems, much larger than can be achieved for instance for the
Ammann-Beenker tiling, because only the one-dimensional Hamiltonian
needs to be diagonalised.

\begin{vchfigure}[tb]
\centerline{\includegraphics[width=0.9\textwidth]{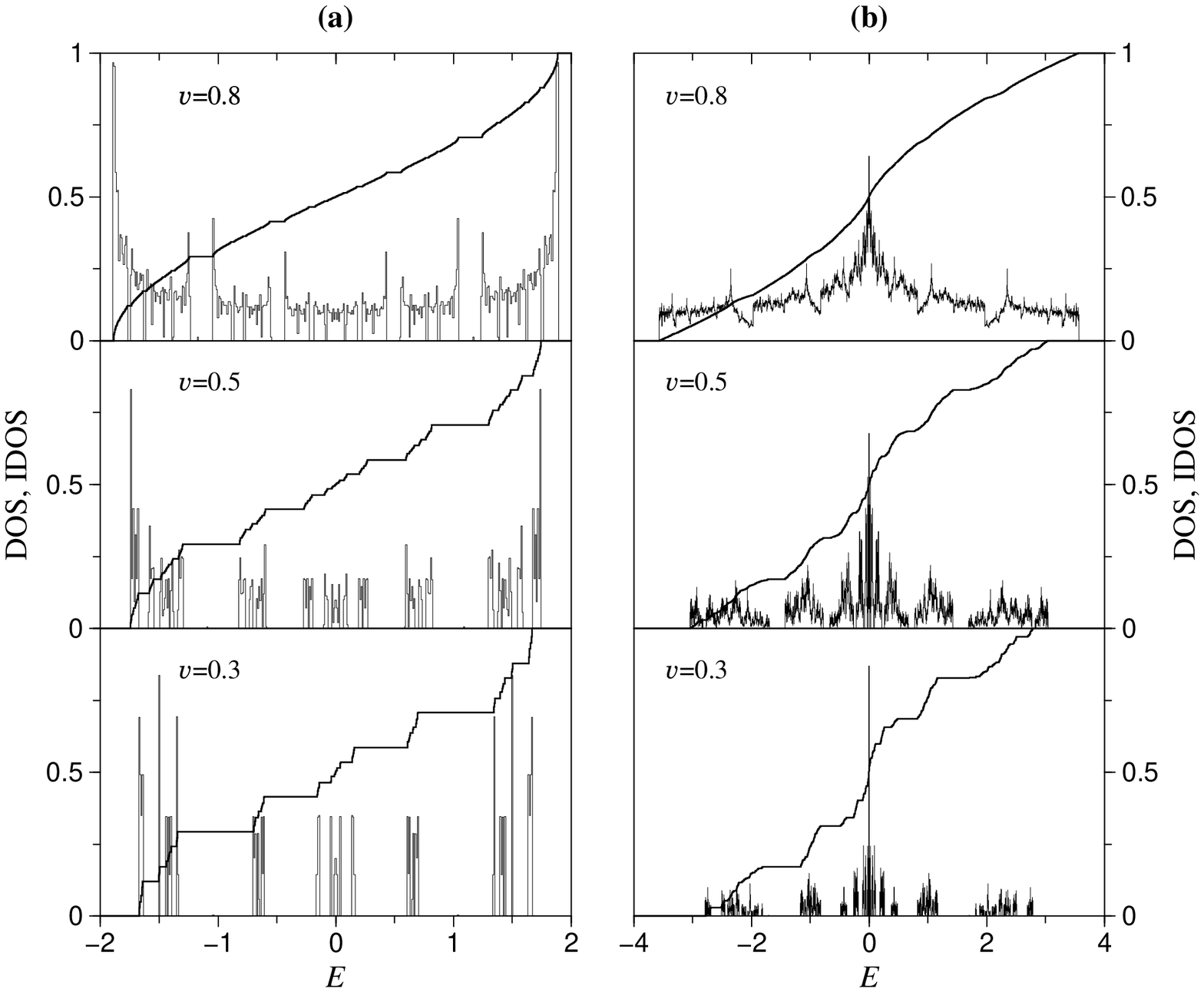}}
\vchcaption{DOS and IDOS for the octonacci chain (left) and the
labyrinth tiling (right).} \label{fig:oldos}
\end{vchfigure}

In Fig.~\ref{fig:oldos}, the DOS and the IDOS for the octonacci chain
and the labyrinth tiling are shown. It appears that for larger values
of $v$ the spectrum of the labyrinth differs qualitatively from that
for small values of $v$, where it splits into many disjoint parts,
similarly to the one-dimensional case. A more detailed investigation
shows that the transition takes place around a value of $v\approx 0.6$
\cite{YGRS00,S89}.

Fig.~\ref{fig:olc} shows numerical results obtained for the temporal
autocorrelation function $C(t)$. Here, we used an octonacci chain of
length $N=19\, 602$, and the corresponding labyrinth thus has
$N^2/2=19\, 602^{2}/2=192\, 119\, 202$ vertices. A qualitative
difference between the behaviour for the one-dimensional and the
two-dimensional system is evident. Whereas in the one-dimensional case
the exponent $\delta$ always changes with $v$, giving $0<\delta<1$ for
all aperiodic systems, this does not seem to be the case for the
two-dimensional system, where $\delta=1$ for values of $v$ between
about $0.6$ and $1$, and $0<\delta<1$ only for small values of
$v$. This is consistent with the observed qualitative change in the
spectrum, as the exponent $\delta$ equals the correlation dimension
$D_{2}$ of the local spectral measure associated with the initial site
\cite{KPG92,ZM95}.\index{spectrum!measure}

\begin{vchfigure}[tb]
\centerline{\includegraphics[width=0.48\textwidth]{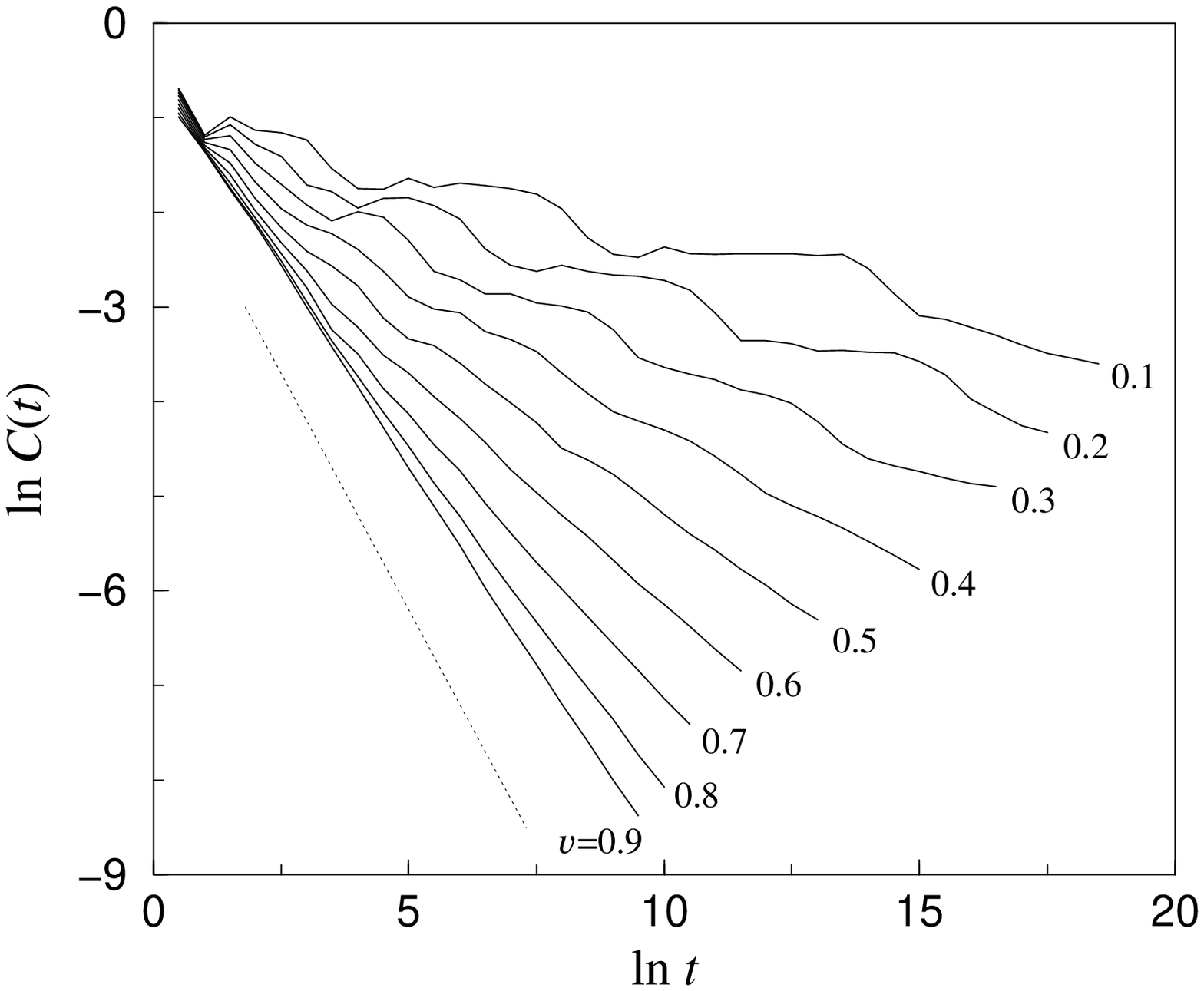}
\hspace*{0.02\textwidth}
\includegraphics[width=0.48\textwidth]{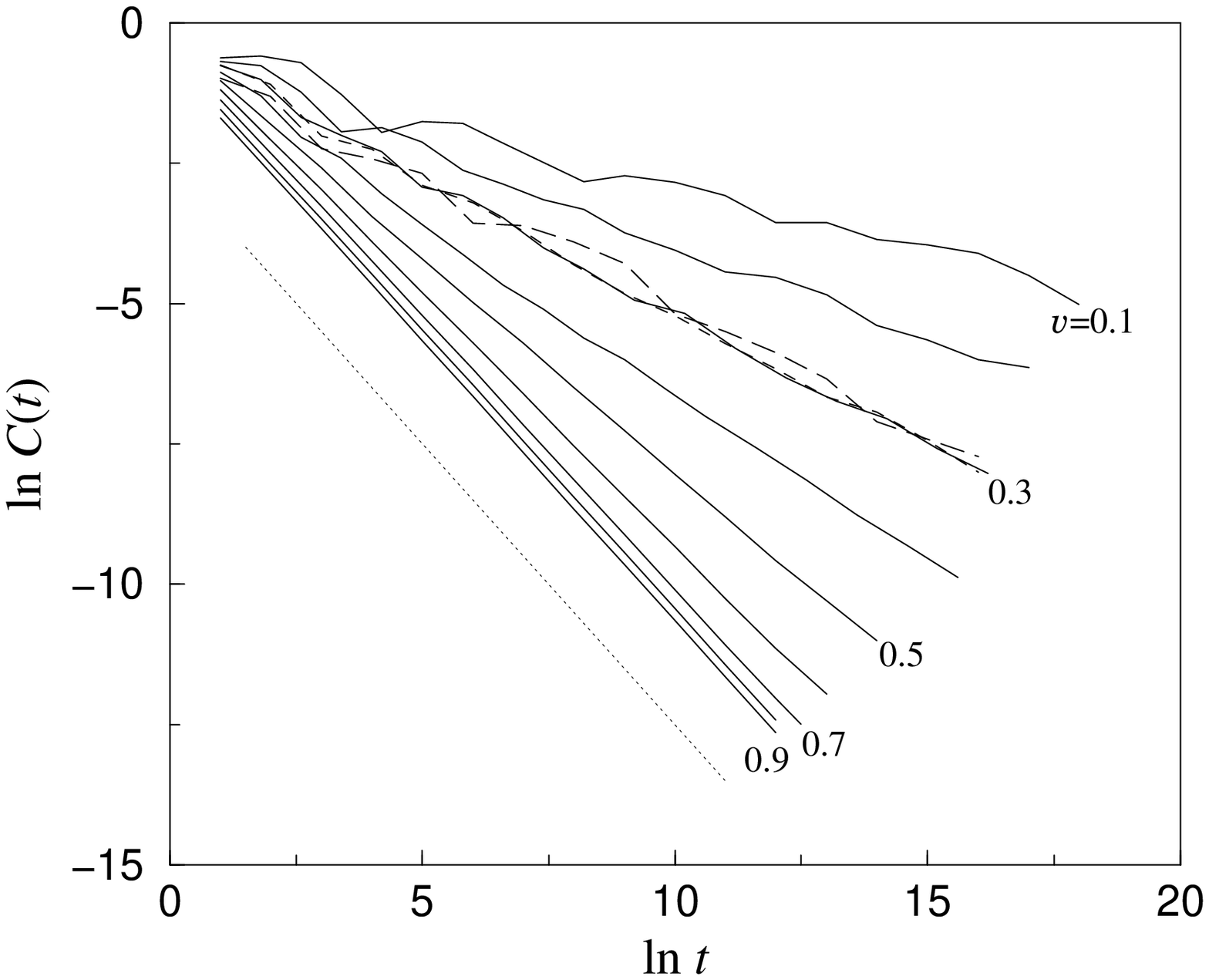}}
\vchcaption{Autocorrelation function $C(t)$ (\ref{eq:cd}) for the
octonacci chain (left) and the labyrinth tiling (right). The
dotted lines correspond to $C(t)\sim t^{-1}$. The dashed lines for
$v=0.3$ correspond to different choices of the initial site
$j_0$.} \label{fig:olc}
\end{vchfigure}

The behaviour of the mean-square displacement is shown in
Fig.~\ref{fig:old}. Here, the system size is smaller, it is $1394$ for
the octonacci chain shown on the left and $578^{2}/2=167\, 042$ for
the labyrinth tiling displayed on the right. In contrast to the
autocorrelation, there is no apparent difference in the behaviour for
the mean-square displacement in the two cases. In particular, there is
no qualitative change around the value $v\approx 0.6$ where we found a
transition in the spectral measure and the autocorrelation function.

\begin{vchfigure}[tb]
\centerline{\includegraphics[width=0.48\textwidth]{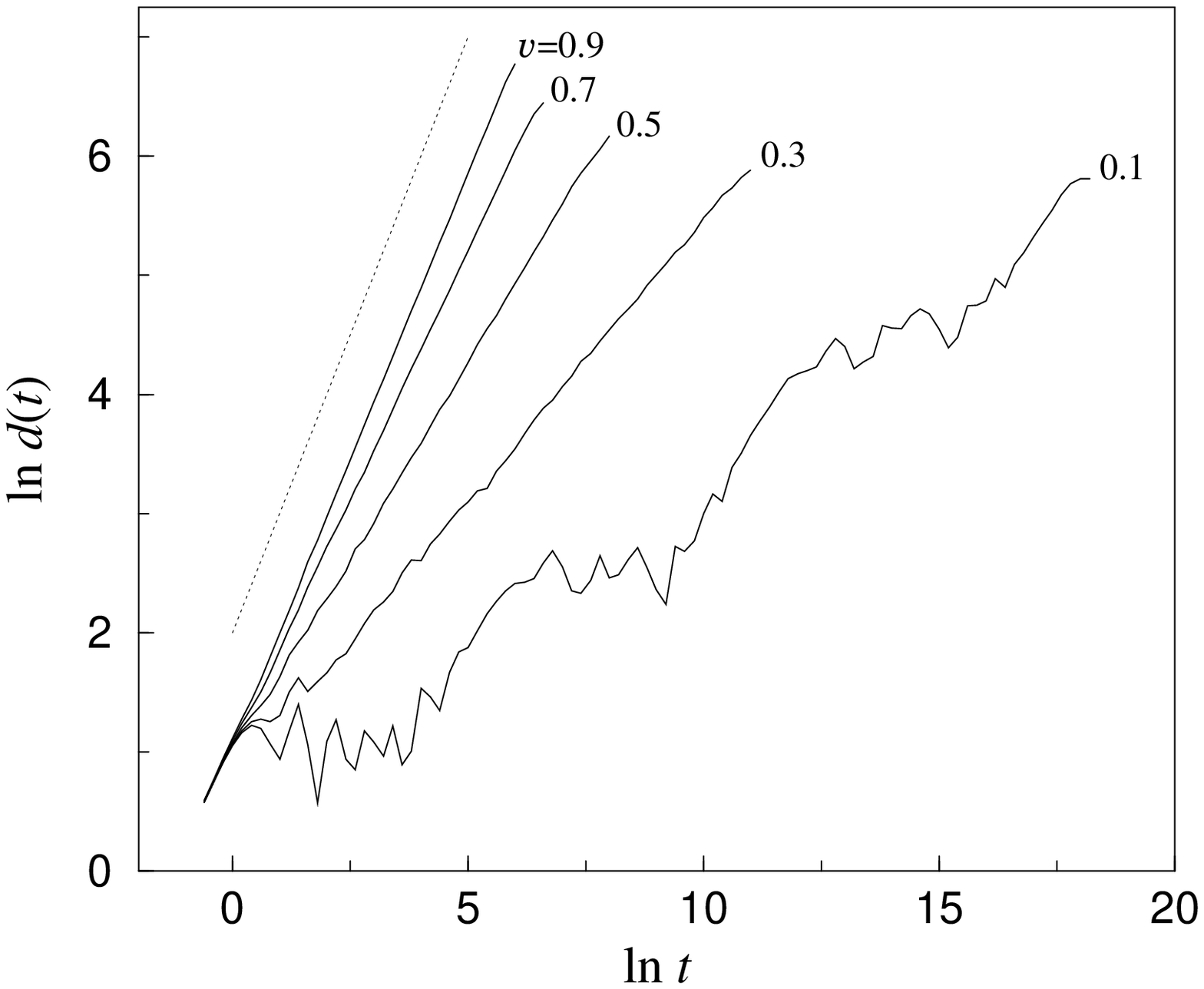}
\hspace*{0.02\textwidth}
\includegraphics[width=0.48\textwidth]{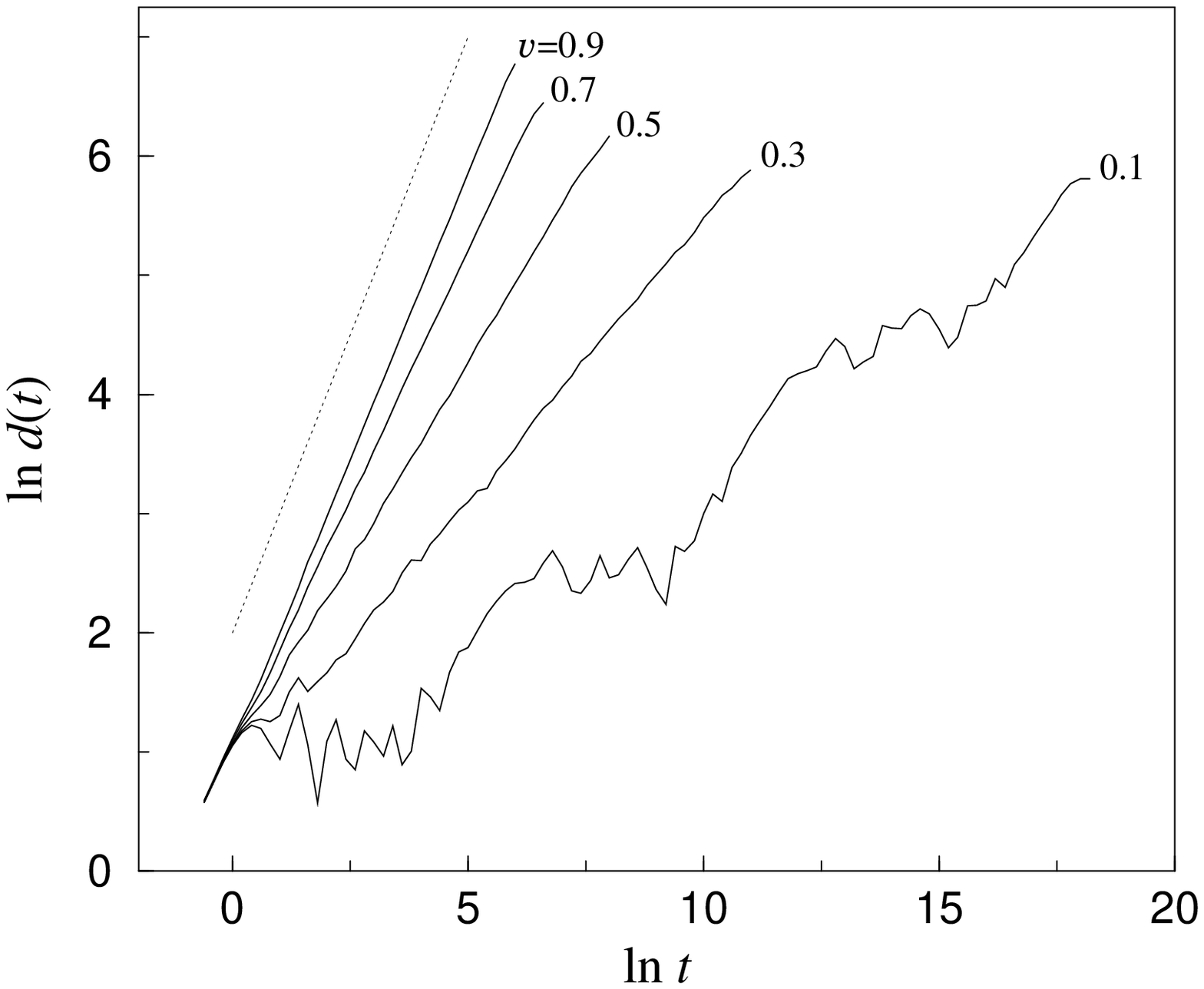}}
\vchcaption{Mean-square displacement $d(t)$ (\ref{eq:cd}) for the
octonacci chain (left) and the labyrinth tiling (right). The
dotted lines corresponds to ballistic motion $d(t)\sim t$.}
\label{fig:old}
\end{vchfigure}

\begin{vchfigure}[tb]
\centerline{\parbox[b]{0.5\textwidth}{%
\includegraphics[width=0.5\textwidth]{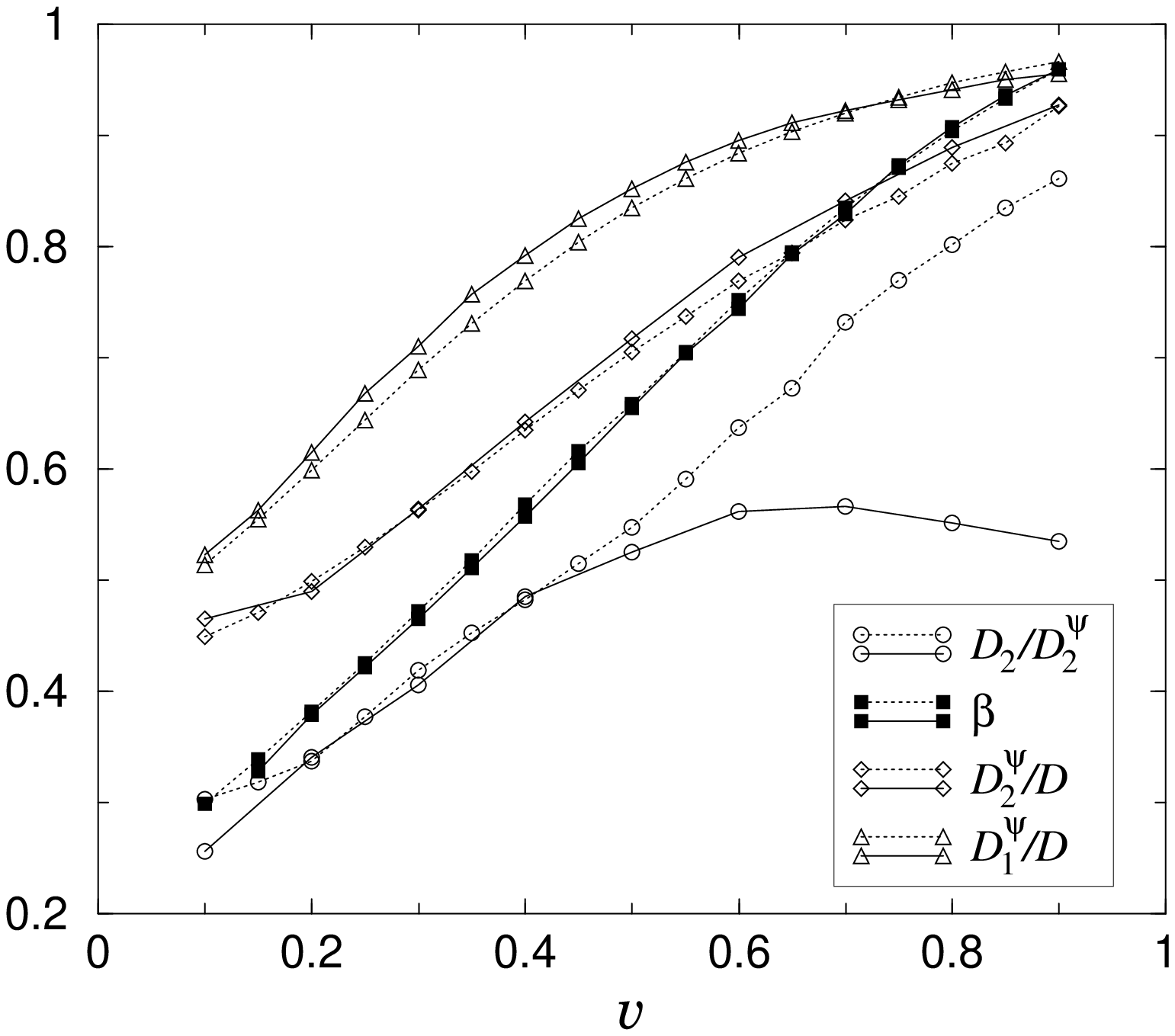}}%
\parbox[b]{0.5\textwidth}{%
\vchcaption{The exponent $\beta$ compared to several quantities
describing the multifractal properties of the energy spectra and
wavefunctions. The lines are included to guide the eye; the dotted
lines correspond to the results for the octonacci chain (dimension
$D=1$), the solid lines to the labyrinth tiling ($D=2$).\bigskip}
\label{fig:dim}}}
\end{vchfigure}

Several inequalities and approximations have been derived that relate
spectral properties with the exponent $\beta$ characterising quantum
diffusion. Quasiperiodic tight-binding Hamiltonians, due to their
intricate spectral properties, provide a particularly challenging test
to these results. In Fig.~\ref{fig:dim}, the numerical values for
$\beta$ for the octonacci chain and the labyrinth tiling, for various
values of the hopping parameter $v$, are compared with expressions
involving the fractal dimensions\index{fractal dimension} $D_{q}$ of
the spectral measure\index{spectrum!measure} and $D^{\psi}_{q}$ of the
eigenstates. In particular, the bound $\beta \geq D_{2}/D_{2}^{\psi}$
\cite{KKKG97} is numerically obeyed by both the octonacci chain and
the labyrinth tiling. However, the inequality is less sharp in the
latter case as $\beta$ is much larger than the ratio
$D_{2}/D_{2}^{\psi}$, in particular for values of the parameter $v\ge
0.6$ where the energy spectrum is smooth and $D_{2}\approx 1$. Another
expression \cite{ZZSPN}, $\beta\le D_{2}^{\psi}/D$, where $D$ denotes
the spatial dimension, appears to be violated for parameter values of
$v$ close to one, which means close to the periodic case, though it
stays reasonably close to $\beta$ for all values of $v$.  We note that
the weaker condition $\beta\le D_1^{\psi}/D$ appears to be always
satisfied, although we presently do not know of any argument why this
bound should hold. A more thorough understanding of the relations
between the spectral properties and quantum diffusion is clearly
desirable.\index{quantum diffusion}

\subsection{Interacting Electrons}\index{electrons!interacting}
\label{sec:ie}

All previously mentioned results, and indeed most results in the
literature, are based on models of non-interacting electrons. Whereas
the main effects of an interaction can often be accounted for by
considering quasiparticles with effective parameters instead of bare
electrons, it is not clear that electrons in quasicrystals can really
be described in this way. Therefore, it is interesting to study the
effect of an electron-electron interaction on the spectral properties
of aperiodic Schr\"{o}dinger operators. In this case, we choose the
Aubry-Andr\'{e} or Harper model mentioned above,\index{Harper model}
because by changing the strength of the aperiodic modulation we can
investigate extended states, critical states and localised states, and
indeed study the effects of an interaction on the metal-insulator
transition.\index{states!critical}\index{states!localised}
\index{metal-insulator transition}

Our Hamiltonian of $N$ interacting spinless fermions on a
ring of circumference $M$ is
\begin{equation}
H=-\sum_{j=1}^{M}c^{\dagger}_{j+1}c^{}_{j}+c^{\dagger}_{j}c^{}_{j+1}
+V\sum_{j=1}^{M}n^{}_{j+1}n^{}_{j}+2\mu\sum_{j=1}^{M}\cos(\alpha
j+\beta) n^{}_{j}, \label{eq:haa}
\end{equation}
where $c^{\dagger}_{j}$ and $c^{}_{j}$ are fermionic creation and
annihilation operators, respectively, and
$n_{j}=c^{\dagger}_{j}c^{}_{j}$ is the corresponding number
operator. The hopping parameter was chosen to be one, and $V$ denotes
the strength of the interaction between the fermions. The parameter
$\mu$ controls the strength of the aperiodic modulation. The
aperiodicity is determined by an irrational number $\alpha/2\pi$ which
we choose as $\alpha/2\pi=1/\tau=(\sqrt{5}-1)/2$, and $\beta$ is an
arbitrary shift.

First, the case of very low density, meaning just two particles with
opposite spin and an on-site (Hubbard) interaction $U$, has been
investigated by means of the transfer-matrix method and finite-size
scaling \cite{EGRS99,ERS01,ERS02}. In Fig.~\ref{fig:scal}, the
behaviour of the reduced localisation lengths
$\Lambda_{M}=\lambda_{M}/M$ \cite{ERS01,ERS02} as a function of $\mu$
is shown for the non-interacting case and the interacting case for
various system sizes $M$. Here $\lambda_{M}$ denotes the localisation
length in the finite samples. The crossing of the interpolating lines
for different $M$ indicates the metal-insulator
transition. Finite-site scaling then yields the correlation lengths
for the infinite systems.  For two interacting particles, the
conclusion is that the metal-insulator transition remains unaffected
by an on-site interaction. The observed critical potential strength is
consistent with $\mu_{c}=1$ also in the interacting case, and the
critical exponent of the correlation lengths $\xi$ is $\nu\approx 1$
\cite{EGRS99,ERS01,ERS02}. For a long-range interaction the accuracy
of the data is not very good, but it was observed \cite {EGRS99} that
the metal-insulator transition tends to shift to smaller values of the
critical potential strength $\mu_{c}\approx 0.92$.

\begin{vchfigure}[tb]
\centerline{\includegraphics[width=0.5\textwidth]{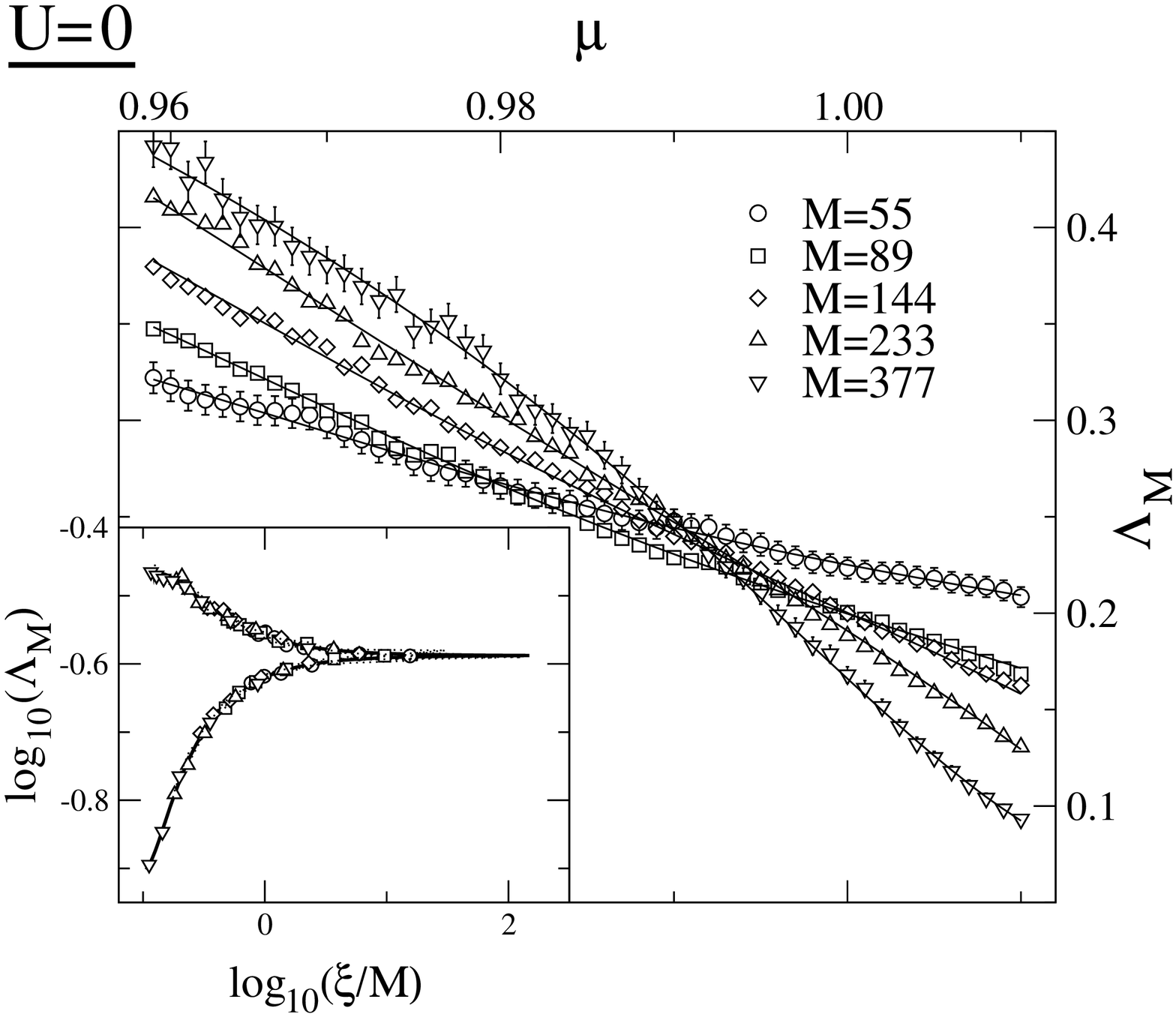}%
\includegraphics[width=0.5\textwidth]{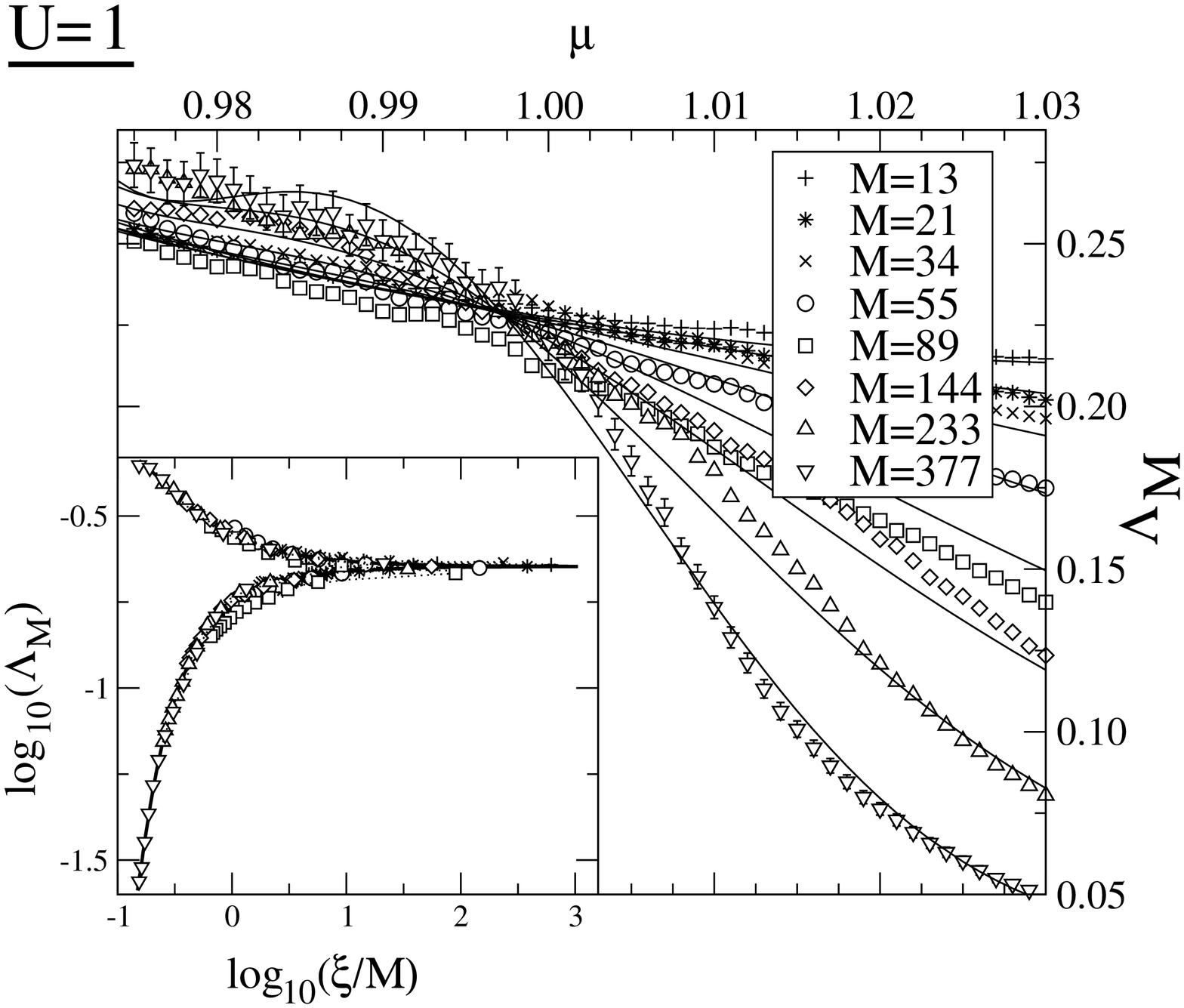}}
\vchcaption{Dependence of the reduced localisation lengths
$\Lambda_{M}$ for systems of size $M$ on the strength $\mu$ of the
aperiodic potential, without interaction, $U=0$ (left), and for
two interacting particles with $U=1$ (right). The insets show the
scaling function in dependence on the correlation length $\xi$.}
\label{fig:scal}
\end{vchfigure}

More interesting is the case of finite density $\varrho=N/M$. This has
been investigated by the density-matrix renormalisation-group method
\cite{SRS02a,SRS02b} which allows us to treat systems of length up to
$M\approx 100-200$, whereas direct diagonalisation techniques are
limited to very small system sizes. A convenient observable to
investigate the metal-insulator transition, in this case, is the phase
sensitivity
\begin{equation}
M\, \Delta E = (-1)^{N}\, M\, [E(0)-E(\pi)]
\label{eq:ps}
\end{equation}
of the ground state \cite{SRS02a,SRS02b}, where $E(\varphi)$ denotes
the ground-state energy of the system with a twist
$c^{}_{M+1}=\exp(i\varphi)c^{}_{1}$ at the boundary. This quantity is
independent of the system size for extended wave functions and
decreases exponentially with system size if the wave function is
localised.  In this case, it turns out that the behaviour of the
system depends on the particle density $\varrho$. If $\varrho$ is
incommensurate with the parameter $\alpha/2\pi$, the system behaves
very much like the one with two interacting particles. For attractive
or repulsive interaction, one finds a critical potential strength
$\mu_{\rm c}\approx 1$ and a critical exponent $\nu\approx 1$ for the
phase sensitivity $M\, \Delta E\sim (\mu_{\rm c}-\mu)^{\nu}$ near
criticality.

For particle densities $\varrho$ that are commensurate with
$\alpha/2\pi=\tau^{-1}$, for instance for $\varrho_{n}=\tau^{-n}$, a
different behaviour is observed. This may be understood most easily
for $\varrho_{1}=\tau^{-1}$ where the resonance condition
$\alpha=2k_{\rm F}$ holds for the Fermi wave vector $k_{\rm
F}=\pi\varrho$. In this case, one may expect to find a Peierls-like
transition \cite{Pei55}, and this is indeed observed
\cite{SRS02a,SRS02b}. The phase diagram for fixed commensurate density
$\varrho$ appears to be dominated by localised states, at a given
potential strength $\mu>\mu_{\rm min}$, with a regime of extended
states for small values of $\mu$ in a certain range of interaction
strengths $V$. For repulsive and weakly attractive interactions, the
ground state for $\mu>\mu_{\rm min}$ is localised. For strongly
attractive interactions, the system shows Peierls-like
behaviour,\index{Peierls transition} meaning that there is a
transition from the insulating to a metallic phase around
$V\approx-\sqrt{2}$ \cite{SRS02a,SRS02b}. As an example, the scaling
function obtained for the density $\varrho_{3}=\tau^{-3}\approx 0.236$
and potential $\mu=0.4$ is shown in Fig.~\ref{fig:scal3}.

\begin{vchfigure}[tb]
\centerline{\parbox[b]{0.6\textwidth}{%
\includegraphics[width=0.6\textwidth]{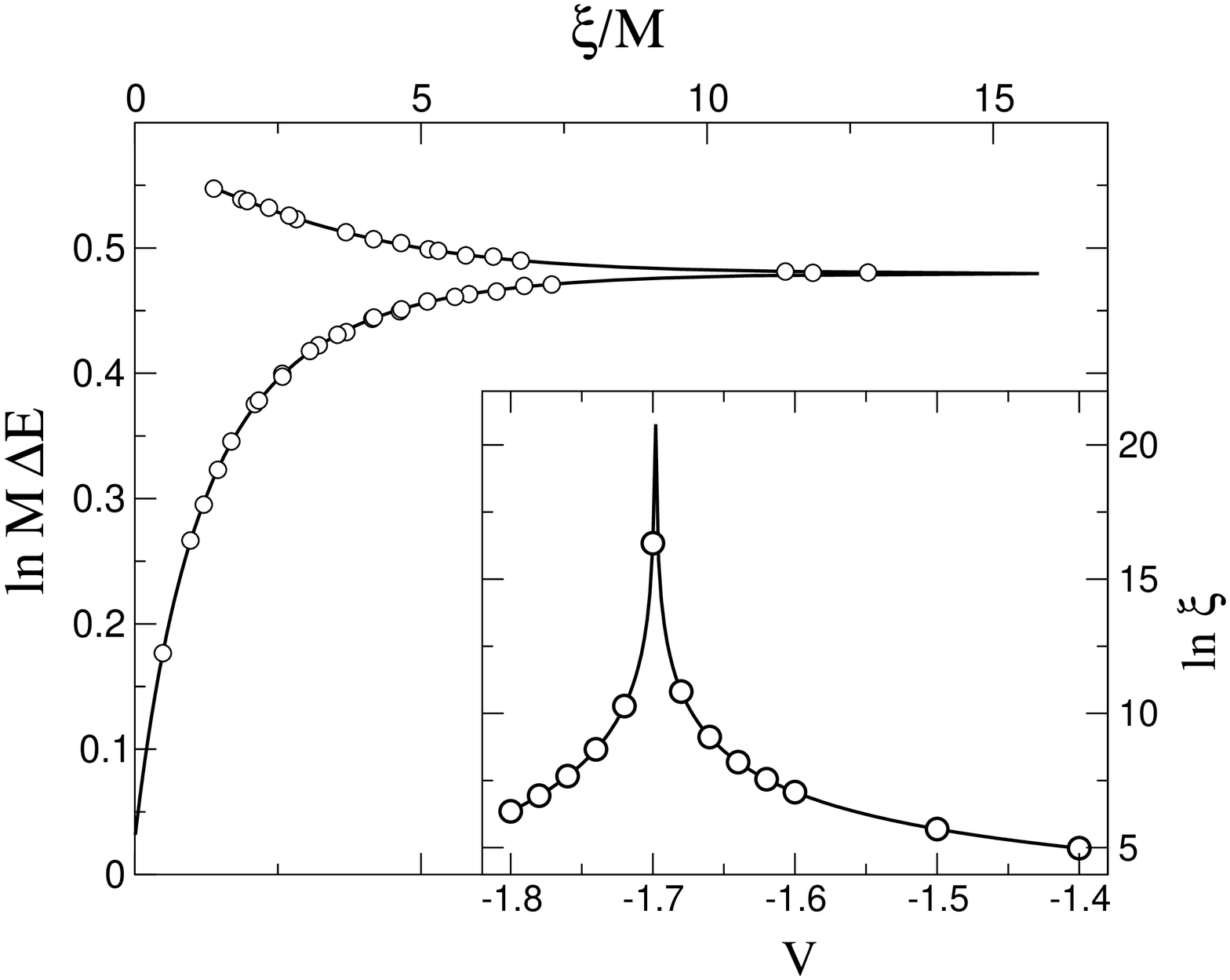}}%
\parbox[b]{0.4\textwidth}{%
\vchcaption{Scaled data of the phase sensitivity $M\, \Delta E$
(\ref{eq:ps})
for the model of Eq.~(\ref{eq:haa}) with parameter
$\alpha/2\pi=\tau^{-1}$, particle density $\varrho_{3}=\tau^{-3}$,
potential $\mu=0.4$, interaction parameters $V\in[-1.8,-1.4]$
and various system sizes $M$. The inset shows the scaling
parameter $\xi$.\bigskip}\label{fig:scal3}}}
\end{vchfigure}

In summary, for the incommensurate case the interaction does not seem
to affect the metal-insulator transition in this system. For the
commensurate case, the behaviour is dominated by the resonance which
leads to Peierls-type behaviour.

\subsection{Concluding Remarks}

Though there has been considerable progress over the last two decades,
our understanding of the physical properties of quasicrystals is still
far from complete, as is the knowledge about their structure
\cite{S99,SSH02} and the physical growth mechanism
\cite{GJ02}. Concerning their electronic properties, there exist
several approaches that can, at least, qualitatively account for the
peculiar features of quasicrystals.

Here, we discussed spectral properties of discrete aperiodic
Schr\"{o}dinger operators based on numerical and analytic
investigations. Whereas one-dimensional models are reasonably well
understood, our present knowledge of the two-dimensional systems rests
largely on numerical investigations of a few examples, notably the
Penrose and Ammann-Beenker tilings. For these models, numerical
results establish multifractal behaviour of generic eigenstates, which
is corroborated by the analytical construction of such states for the
Penrose tiling.  A statistical investigation of the energy spectrum
shows, somewhat surprisingly, that the level-spacing distribution of
such models appears to be extremely well described by universal random
matrix distributions, and that there is no ``critical'' statistics as
one might have guessed from the multifractal character of the
eigenstates.  Furthermore, we discussed quantum diffusion by
numerically investigating a tight-binding model on a peculiar tiling,
the labyrinth, which has the property that all eigenenergies and
eigenstates can be obtained as products of those of a one-dimensional
tight-binding model on the octonacci chain. Our results corroborate
that there are strong relations between fractal properties of energy
spectra and wavefunctions on the one hand and the exponents describing
the quantum diffusion on the other hand. However, it appears to be
difficult to find relations that give quantitative agreement for one-
and two-dimensional aperiodic systems. Here, a deeper understanding of
the underlying physics is desirable. Higher-dimensional systems
constructed as products of one-dimensional systems, such as the
labyrinth tiling, may provide useful toy examples for further
investigations which can, at least, be treated numerically in an
efficient way.  Finally, the role of an electron-electron interaction
on the metal-insulator transition in a one-dimensional aperiodic
system was investigated. The results show that resonance-type
phenomena are important, giving rise to a different behaviour for
particle densities which are commensurate and incommensurate with the
modulation. In the incommensurate case, only minor effects can be
seen, whereas Peierls-like transitions are observed for the
commensurate case.

Not much is known for three-dimensional systems, though it appears
plausible, and consistent with our numerical results, that there is a
general tendency that eigenstates become less localised with
increasing spatial dimension. However, many questions remain
unanswered, and it is not even decided whether extended states exist
in models such as the tight-binding model on the Ammann-Kramer-Neri
tiling considered here.  Other interesting questions concern transport
in quasiperiodic systems in the presence of magnetic fields, see for
instance \cite{GGS98} for a glimpse at the complexity of such systems.

\section*{Acknowledgement}

This work was supported by Deutsche Forschungsgemeinschaft (Schr
231/16).  UG would like to thank the Erwin Schr\"{o}dinger
International Institute for Mathematical Physics in Vienna for support
during a stay in December 2002 where this work was completed.


\end{document}